\documentclass{JHEP}
\usepackage{epsfig}
\usepackage{amssymb,latexsym,amsfonts}

\newcommand{\be}{\begin{equation}}
\newcommand{\ee}{\end{equation}}
\newcommand{\bd}{\begin{displaymath}}
\newcommand{\ed}{\end{displaymath}}
\newcommand{\ba}{\begin{array}}
\newcommand{\ea}{\end{array}}
\newcommand{\bt}{\begin{tabular}}
\newcommand{\et}{\end{tabular}}

\newcommand{\bea}{\begin{eqnarray}}
\newcommand{\eea}{\end{eqnarray}}
\newcommand{\bean}{\begin{eqnarray*}}
\newcommand{\eean}{\end{eqnarray*}}
\newcommand{\non}{\nonumber}

\newcommand{\hlf}{\frac{1}{2}}
\newcommand{\qrt}{\frac{1}{4}}
\newcommand{\tqrt}{\frac{3}{4}}
\newcommand{\thlf}{\frac{3}{2}}

\newcommand{\dif}{\mathrm{d}}

\newcommand{\inp}[2]{\langle #1, #2 \rangle}

\newcommand{\Z}{\mathbb{Z}}

\newcommand{\R}{\mathbb{R}}

\preprint{{\tt hep-th/0607069}}
\title{Determining the dual}

\author{Arjan Keurentjes
\\ {\it Theoretische Natuurkunde, Vrije Universiteit Brussel, and\\
 the International Solvay Institutes,\\Pleinlaan 2, B-1050 Brussels, Belgium}
\\ \email{arjan@tena4.vub.ac.be} }

\abstract{We study the $R \rightarrow 0$ limit for heterotic strings of either kind ($Spin(32)/\Z_2$ or $E_8 \times E_8$) compactified on a circle, in the presence of an arbitrary Wilson line. Though for generic Wilson line this limit leads to chaotic behaviour, there are two distinguished, countable subsets of Wilson lines, that are dense in the total space of Wilson lines: One subset leads to decompactification limits; a second subset converges onto periodic orbits. Many of the implications carry over to heterotic strings on a circle of small but finite radius. To complete the picture, we discuss global aspects of the moduli-space, compare it with the ``fiducial'' moduli-space for type I strings on a circle, give a derivation of the map between the moduli of the two heterotic string theories on a circle at an arbitrary point in the moduli space, and compute the smallest radius that can be probed.}  

\begin{document}
\section{Introduction}

Duality transformations and symmetries have become commonplace in the high energy physics literature. In regimes where physical computations seem difficult, duality transformations provide a map to other regimes, or even other models where computations may be done with more ease. It has been known however that duality cannot always help us, as duality transformations may take us to theories or descriptions that do not allow a simple analysis, for example because it is not clear how to describe the theory at weak coupling(relatively ``simple'' examples can be found in \cite{Vafa:1996xn, Cachazo:2000ey}).

In the cases where one is in familiar territory, T-dualities map small volumes to large volumes, S-dualities map strong coupling regimes to weak coupling regimes. The closed bosonic string on a circle of radius $R$ is T-dual to the closed bosonic string on a circle of radius $1/R$. Slightly more involved but still simple, the IIA-string on a circle is T-dual to the IIB string on a circle of inverse radius. Perhaps the simplest example of an S-duality is the relation between IIA string theory and 11-dimensional supergravity/M-theory, where strong coupling in the IIA string should lead to a description in terms of 11-dimensional supergravity on a large circle.

In these examples there is only one modulus involved (in all of the above examples, the radius of a circle). For the heterotic string on a circle however, there are 17 moduli, mixing under duality, and things are (much) more intricate. In the $Spin(32)/\Z_2$ theory, turning on the Wilson line
\be
A_S = \omega_S = \left( \left(\hlf \right)^8 , 0^8 \right)
\ee
breaks the gauge symmetry to a group with algebra $so(16) \oplus so(16)$ (the actual group is actually $(Spin(16) \times Spin(16))/\Z_2$ \cite{deBoer:2001px}, and we are supressing $U(1) \times U(1)$ due to the graviphoton and the dimensionally reduced $B$-field). This theory at radius $R$ is dual to the $E_8 \times E_8$-theory at radius $1/R$ \cite{Ginsparg:1986bx}, with a Wilson line given by
\be
A_E =\omega_E = \left( 0^7, 1, -1, 0^7 \right).
\ee
This is an elegant result. It also fits well with the naive picture of the strong coupling limits of the heterotic string theories: The $E_8 \times E_8$ heterotic string is dual to M-theory on an interval \cite{Horava:1995qa}, or after a 9-11 flip to type IIA on an interval (also known as type I'); the $Spin(32)/\Z_2$ heterotic string is dual to type I, which is an orientifold of type IIB; Type IIA and type IIB, as well as their orientifolded couterparts type I' resp. Type I are T-dual to each other.

Unfortunately this picture is too naive, because the presence of the Wilson lines is crucial for this duality. As already observed by Ginsparg and others \cite{Ginsparg:1986bx}, when we set the Wilson line to $A_S = 0$, then the $Spin(32)/\Z_2$-theory at radius $R$ is dual to the $Spin(32)/\Z_2$-theory at radius $2/R$ (with $A_S=0$). A similar result is true for the $E_8 \times E_8$-theory; compactifying it on a circle with radius $R$ and Wilson line $A_E = 0$, it is dual to $E_8 \times E_8$-theory on a circle of radius $2/R$ with $A_E=0$. 

The 2 choices for $A_S$ and $A_E$ are no more than 2 points in the continuum of possible Wilson lines, and a natural question is what happens for arbitrary Wilson line at small radius. We should specify also what we mean by ``small radius'' (in view of the different coefficients in the above dualities). At first we will adopt a naive point of view, and attempt to chase the theory to the extreme limit $R \rightarrow 0$ (at fixed Wilson line). 

It is easily seen that this corresponds to tracking down along a geodesic in the moduli space, which is locally of the form $SO(17)\backslash SO(17,1)$. The global properties are determined by the action of a discrete group $O(\Pi_{17,1})$ \footnote{This group is also known under other names. $O(\Pi_{17,1})$ is a discrete subgroup of $O(17,1)$ consisting of transformations that map the self-dual lorentzian lattice $\Pi_{17,1}$ to itself. We give its generators in section \ref{Ogen}.} acting from the right. The generators for this group are easy to find, but it is not trivial to study the action of the full group. Moving along generic geodesics on this space, taking care to map all the segments of the geodesic to a suitably chosen fundamental domain, is (well-)known to lead to chaotic behaviour \cite{Horne:1994mi}. There is however a countable subset of (infinitely many) Wilson lines that do not lead us into chaos. Analyzing these will be the most important topic of this paper.

To develop intuition, we study the much simpler case of $SL(2,\Z)$-duality, where the explicit action and group are well-known. Section \ref{SL2Z} is devoted to this group, and prepares for the methods that we will use on the heterotic string.

In section \ref{OPi171} we will turn to heterotic strings on a circle, and describe the moduli-space for these theories. We also, briefly, return to the issue that dual descriptions in terms of the type I and type I' theories do not seem to cover all corners of the full moduli-space.

Subsequently we study the limit $R \rightarrow 0$. We distinguish 4 possible outcomes:
\begin{enumerate}
\item The $R \rightarrow 0$ limit corresponds to decompactifaction to the $E_8 \times E_8$-theory;
\item The $R \rightarrow 0$ limit corresponds to decompactifaction to the $Spin(32)$-theory;
\item The limit $R \rightarrow 0$ converges onto a periodic orbit in the moduli space;
\item The limit $R \rightarrow 0$ corresponds to a chaotic orbit in the moduli-space,
\end{enumerate}
and analyze the conditions that lead to the different behaviours. 

The fundamental domain still contains points where the radius of the theory can become very small. Such descriptions with small radius can still be avoided, because there are 2 sets of coordinates covering the fundamental domain, one appropriate for $E_8 \times E_8$-theory, and one for the $Spin(32)/\Z_2$-theory. We give the radius and Wilson line in the dual description as functions of the radius and Wilson line in the original theory. This version of the map is simple, explicit, and symmetric between the two different heterotic theories. As an aside, the map implies that the smallest radius that can be probed by a heterotic string actually depends on the Wilson line, and we compute the minimal value that the radius can take. Subsequently we pay attention to a phenomenon that is responsible for some interesting effects: If both the $R \rightarrow \infty$ and $R \rightarrow 0$ limit give rise to an $E_8 \times E_8$-theory, the assembly of the two $E_8$-factors that takes place in these two limits is not necessarily identical.

We wrap up with a discussion and end with our conclusions.

\section{Warming-up: $SL(2,\Z)$-duality} \label{SL2Z}

Before turning to heterotic strings, we first consider a system realizing an $SL(2,\Z)$ duality. Such systems have two moduli, say a dilaton and an axion. We study the extreme values of the dilaton while keeping the axion fixed. With the explicit form of the $SL(2,\Z)$-transformations, the computations are straightforward and simple. Reanalyzing the result from a different perspective, we will develop methods that generalize to the heterotic string. Indeed, many aspects of what we will find for $SL(2,Z)$ are qualitatively similar to what we will find for the heterotic string on a circle.

\subsection{Strong coupling at fixed axion}

There are many examples of $SL(2,\Z)$-duality. Modular invariance in conformal field theory, S-duality in gauge theories, and the conjectured exact symmetry of the non-perturbative extension of type IIB-string theory (a.k.a. $F$-theory) all feature this symmetry. In the geometric setting $SL(2,\Z)$ is realized as the mapping class group of a 2-torus or elliptic curve. In string theory all the above examples are mixed with and sometimes imply each other.

The most familiar realization of $SL(2,\Z)$ acts on a complex parameter $\tau$. The parameter $\tau$ consists of a real part that is an arbitrary real number, and an imaginary part that is large than 0. Hence $\tau$ parametrizes the upper half plane.

Representing $SL(2,\Z)$ by $2 \times 2$ matrices with integer entries, the group is generated by
\be
S = \left( \ba{rr} 0 & -1 \\ 1 & 0 \ea \right), \qquad
T = \left( \ba{rr} 1 & 1  \\ 0 & 1 \ea \right).
\ee
The action of $SL(2,\Z)$ on the upper half-plane is given by 
\be \label{SL2z}
\tau \rightarrow \frac{a\tau + b}{c\tau +d} \qquad ad-bc=1
\ee
This action is insensitive to the element $-1$ of $SL(2,\Z)$, and therefore the actual group is $PSL(2,\Z) = SL(2,\Z)/\Z_2$ (but typically additional fields break this $\Z_2$ restoring the symmetry to $SL(2,\Z)$). For concreteness, we imagine that this $SL(2,\Z)$ is realized in a gauge theory, or in type IIB-theory. We now want to investigate the strong coupling limit at fixed theta-angle/axion. As the weak coupling limit of these theories corresponds to $\textrm{Im}(\tau) \rightarrow \infty$, this amounts to sending $\textrm{Im}(\tau) \rightarrow 0$ while keeping $\textrm{Re}(\tau))$ fixed. 

The question is now whether there exists an $SL(2,\Z)$ transformation that maps us back to $\textrm{Im}(\tau)\rightarrow \infty$. We first write the action of $SL(2,\Z)$ on $\textrm{Re}(\tau)$ and $\textrm{Im}(\tau)$: 
\be
\textrm{Re}(\tau) + i \textrm{Im}(\tau) \qquad \rightarrow \qquad \frac{(a \textrm{Re}(\tau) + b)(c \textrm{Re}(\tau) + d) + ac (\textrm{Im}(\tau))^2 + i \textrm{Im}(\tau)}{(c \textrm{Re}(\tau) + d)^2 + c^2 (\textrm{Im}(\tau))^2} 
\ee
Mapping the limit $\textrm{Im}(\tau) \rightarrow 0$ to $\textrm{Im}(\tau)\rightarrow \infty$ is only possible if we can eliminate the factor $(c \textrm{Re}(\tau) + d)^2$ from the denominator, in other words if there exist integer $c$ and $d$ such that 
\be
\textrm{Re}(\tau) = -\frac{d}{c}
\ee
In other words: The strong coupling limit $\textrm{Im}(\tau) \rightarrow 0$ can only be mapped to the weak coupling limit $\textrm{Im}(\tau)\rightarrow \infty$ if the theta-angle/axion is \emph{rational}. 

If $\textrm{Re}(\tau)$ is rational, we can and should require that $c$ and $d$ have greatest common divisor $\gcd(c,d) =1$. Via Euclids algorithm, the condition $\gcd(c,d) =1$ implies that a (non-unique) solution to the requirement $ad-bc=1$ exists. 

Hence, for rational $\textrm{Re}(\tau)$, and appropriate $SL(2,Z)$ transformation mapping the limit $\textrm{Im}(\tau)\rightarrow 0$ to $\textrm{Im}(\tau) \rightarrow \infty$ exists, and takes the form
\be
-\frac{d}{c} + i \textrm{Im}(\tau) \qquad \rightarrow \qquad \frac{a}{c} + \frac{i}{c^2 \textrm{Im}(\tau)} 
\ee
This equation suffers from ambiguities in the theta-angle/axion value, on both sides. Shifting the value of $-d/c$ by an integer amounts to replacing $(a,b,c,d)$ by $(a,b+a, c, d+c)$. This preserves $ad-bc=1$ and represents multiplication by $T$ from the right.On the other hand, shifting $a/c$ by an integer amounts to $(a,b,c,d) \rightarrow (a+c,b+d,c,d)$, which represents multiplication by $T$ from the left. These ambiguities can be fixed by demanding that $\tau$ and hence $\textrm{Re}(\tau)$ take values in a suitably chosen fundamental domain, such as the standard choice, the interval $I$ defined by $-\hlf \leq \textrm{Re}(\tau) < \hlf$.

Provided the theta-angle/axion is taking a rational value $-d/c$, the dual value for $\textrm{Re}(\tau)=a/c$. It is immediately obvious that this does not define a function from the interval $I$ to itself, as the function is not defined on the irrational values. Restricting to rational values, the function $-d/c \rightarrow a/c$ is a function, but a very irregular one. Due to the requirement $\gcd(c,d) =1$, $c$ fluctuates wildly when scanning the rational numbers within the interval $I$. And even when we restrict to fixed $c$, then (provided $c$ is large enough) $a$ is a highly irregular function of $d$.

The above considerations do not help us to understand what happens for irrational $\textrm{Re}(\tau)$, hence we turn to other methods.

\subsection{Geodesics}

The group $SL(2,\Z)$ has the action (\ref{SL2z}) on the upper half plane. Writing $\tau = x+ iy$, the Poincar\'e metric on the upper half plane is given by
\be \label{Poin}
\dif s^2 = \frac{\dif x^2 + \dif y^2}{y^2}
\ee
The geodesics for this metric are easy to find. They take the shape of semicircles with midpoint on the (real) $x$-axis, and the two integration constants correspond to the radius, and the midpoint (which, as it lies on a line corresponds to only 1 parameter). The limit case of "infinitely large circles" also makes sense (provided we move the midpoint of the circle such that one of its ends remains fixed), and corresponds to straight lines with $x=c$. 

We were studying the limit $y=\textrm{Im}(\tau) \rightarrow 0$ while keeping $x=\textrm{Re}(\tau)$ constant. This corresponds to moving along a geodesic. Because taking the limit is taking us in a particular direction along the geodesic, it is natural to endow the geodesic with a specific orientation (an arrow).  

We are dealing however, not with the full upper-half plane, but with the upper half-plane quotiented by $PSL(2,\Z)$. It is easily checked that the Poincar\'e metric (\ref{Poin}) is invariant under $PSL(2,\Z)$. A corollary is then that $PSL(2,\Z)$ maps geodesics to geodesics. We should now choose a fundamental domain for this symmetry. The standard choice/picture looks as follows:

\FIGURE{
\includegraphics[width=13cm]{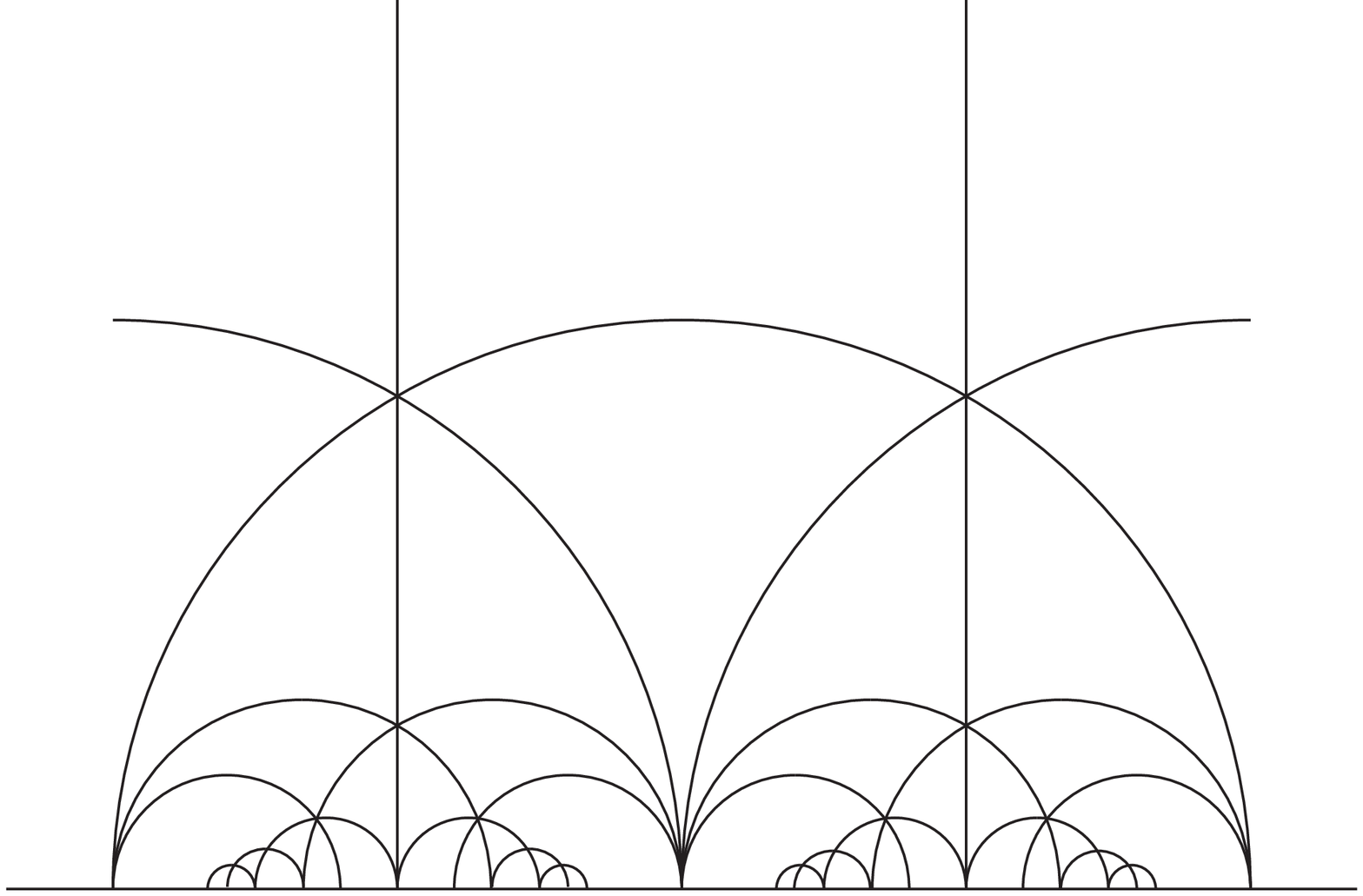}
\caption{Fundamental domains for the $SL(2,\Z)$ action on the upper-half plane.}\label{uphalf}
}

The vertical lines sketched are at $\textrm{Re}(\tau) = \pm \hlf$, the horizontal line is the $\textrm{Im}(\tau) = 0$-axis, and due to graphical (and practical) limitations the circles have not been drawn to arbitrary small size. The standard choice of fundamental domain is given by $x \in I$, $|\tau| > 1$, with the line segment $|\tau| =1$, $-\hlf \leq x \leq 0$ included. 

The lines bounding the fundamental domains are a matter of choice. There are only two, pointlike, singularities in the moduli-space (up to $SL(2,\Z)$-transformations, at $\tau =i$, and at $\tau = -\hlf + \frac{i}{2}\sqrt{3}$, and alternative choices for the boundaries are possible, as long as they contain these singularities (after identification, the two singularities are conical, while the remaining identifications imply a form of periodic boundary conditions). We mention these well-known facts because for the heterotic string to be studied later, the boundaries of the fundamental domain are (almost) all singular hyperplanes, and there is (almost) no arbitary choice. The two singularities correspond to fixed points under $S$ and $ST$, that exist due to 
\be \label{SLrel}
S^2=1; \qquad(ST)^3 = 1.
\ee 
At the singularities there is enhanced symmetry, namely the discrete groups $\Z_2$ generated by $S$, respectively $\Z_3$ generated by $ST$ (2 and 3 are the orders of the discrete symmetry in $PSL(2,\Z)$, in $SL(2,\Z)$ these symmetries are $\Z_4$ and $\Z_6$. In $SL(2,\Z)$ a $\Z_2$ symmetry corresponding to the non-trivial center element $diag(-1,-1)$ is always present, also away from the singularities).

Turning back to our topic of interest, the limit $y \rightarrow 0$ with $x$ fixed takes place on the covering space. The geodesic $x=c$ dives below the circle $|\tau|=1$, and then has to be mapped back to the fundamental domain. In general, the geodesic will cross many more fundamental domains. The question whether it ends up at weak coupling ($y \rightarrow \infty$) is the question whether it hits one of the preimages of $\tau = i \infty$. These pre-images are organized as cusps near the (real) $x$-axis, and as is easy to see, these cusps are at rational points, and conversely, for any rational point there exists an $PSL(2,\Z)$ transformation mapping it to $i \infty$. Geodesics aiming for irrational $\textrm{Im}(\tau)$ will ``miss'' all the rational points at the $x$-axis, by an arbitrary small, but non-zero amount.

\subsection{Begin- and endpoints; continued fractions}

With the picture of the limit we are interested as the chasing down of a geodesic, we can analyze in greater detail than thus far.

Supose we are interested in the $\textrm{Im}(\tau) \rightarrow 0$ limit of a theory with fixed, rational $\textrm{Re}(\tau)$. Using axion shifts, we can assume that $\textrm{Re}(\tau) \in I$. We are hence dealing with a geodesic with beginpoint inside the fundamental domain, while the endpoint is outside. Therefore, if we chase down the geodesic, we cross the line $|\tau| =1$. At that point we use an $S$-transformation to transform to another geodesic. This geodesic is a semi-circle, and at first will go ``up'' (to higher values of $y=\textrm{Im}(\tau)$), but eventually come down again. Because of the circle property, by then we will be outside the fundamental domain, as $x$ will be outside the interval $I$. But this is easily remedied by applying a number of $T$-transformations. That we return to the $x$-axis implies however that we need another $S$ transformation, that will take us to another geodesic. 

This procedure is repeated, until an $S$-transformation maps us to a geodesic that is a straight line running of to infinity. If we reach this geodesic, the dual value for $\textrm{Re}(\tau)$ is easily read off: it is the constant $x$-value of the final geodesic.

We will illustrate this with an example. A geodesic is easily characterized by the real numbers representing the begin- and endpoint. For the special case of geodesics extending to $i \infty$, we will call the begin- or endpoint\footnote{The upper half-plane with its boundary at $\textrm{Im}(\tau) = 0$ and $\tau = i \infty$ included is compact; this is a one-point-compactification. Mapping the upper-half plane to a disk (by a Cayley-transform), all lines running off to infinity pass through (the image of) $i \infty$.} $i \infty$. 

Consider now the fixed axion value $\textrm{Re})(\tau)= -19/52$ and vary $y$. This is a geodesic beginning at $\tau = i \infty$, and ending at $\tau =-19/52$. In the subsequent table the geodesic is chased under $SL(2,\Z)$-transformations. Of the image geodesic, we mention the begin- and endpoint, and the transformation that takes us to the next step.
 
\bd
\ba{|r|r||c|}
\hline
\textrm{begin} & \textrm{end} & \textrm{transf.} \\
\hline
 i \infty & -19/52 & S\\
0         &  52/19 & T^{-3}\\
-3        & -5/19  & S\\
1/3 & 19/5 & T^{-4}\\
-11/3 & -1/5 & S\\
3/11 & 5 & T^{-5}\\
-52/11 & 0 & S \\
11/52 & i \infty & \\
\hline
\ea
\ed

After $ST^{-3}ST^{-4}ST^{-5}S$ we have arrived at a straight geodesic, with $x = 11/52$. This is indeed the solution to 
\bd
ad-bc =1 \qquad \rightarrow \qquad (11 \times 19) - (4 \times 52) =1
\ed
(hence $b=4$). The values can also trivially be read off from
\bd
S T^{-3} S T^{-4} S T^{-5}S =-\left( \ba{rr} 11 & 4\\ 52 & 19 \ea \right). 
\ed

When expressed as fractions, or numbers with a decimal expansion the map from axion to dual axion is a highly irregular one. The above procedure suggests another way of expressing the axion-values, namely as a continued fraction\footnote{Our continued fractions have $-1$ in the numerator instead of $1$ because $S$ takes $\tau \rightarrow - \frac{1}{\tau}$.}:
\be
-\frac{19}{52} = {-1 \over 3- {1 \over 4-{1 \over 5}}} \qquad
\frac{11}{52} = \frac{-1}{-5- \frac{1}{-4-\frac{1}{-3}}} \qquad
\ee 
To abbreviate the continued fraction, and save ourselves from typographical trouble, we will use the following notation:
\bd
[a_1,a_2,a_3] = \frac{-1}{a_1 - \frac{1}{a_2 - \frac{1}{a_3}}}
\ed
For continued fractions of different length, we define recursively
\bd
[a_1] = -\frac{1}{a_1} \qquad \frac{-1}{[a_1,a_2, \ldots, a_n]} =a_1 + [a_2, \ldots, a_n]
\ed 
Hence $-19/52 = [3,4,5]$ and $11/52 = [-5,-4,-3]$. Whereas in other notations the relation between axion and dual axion is obscure and erratic, in this notation it is completely obvious: To obtain the dual axion from a (rational) value of the axion, simply write the axion as a continued fraction as above. The dual axion is then given by the sequence of the axion, ordered backwards, and every entry $a_i$ replaced by $-a_i$. 

Of course it is crucial for this procedure to work that the continued fraction terminates. This automatically implies that we are dealing with a rational number. Taking a real and rational number through the repeated steps of translating to the fundamental domain ($\textrm{Re}(\tau) \rightarrow \textrm{Re}(\tau)\pm 1$ until $\rm{Re}(\tau) \in I$) and inversion ($\textrm{Re}(\tau) \rightarrow -1/\textrm{Re}(\tau)$), the absolute value of the denominator decreases in each step. When the denominator becomes $1$, we translate the total axion to $0$, and then inversion will give us an endpoint at $i \infty$. Hence, any rational number has a terminating continued fraction expansion.

\subsection{Strong coupling for irrational axion}

With irrational axion, one can write down a formal continued fraction, but the algorithm of translation to the fundamental domain and inversion now never terminates. There is also no value for a dual axion; the algorithm for determining it does not work, as it does in fact not exist.

Of course this is far from dramatic; having declared from the outset that we are interested in a limit, there was always the risk that this limit will not converge. Convergence is achieved for the rational values of the axion, and that the rationals are dense in the interval $I$, but irrational numbers never lead to convergence. There is a subclass of the irrationals that leads to something close to convergence. 

An interesting subcategory of the infinite continued fractions have some periodicity in their expansion. The easiest examples of these are the continued fractions
\be
x = [a,a,a,a,a,\ldots]
\ee
These continued fractions obey the equation
\be
-\frac{1}{x} = a + x
\ee
and hence
\be \label{sol}
x = -\frac{a}{2} \pm \frac{\sqrt{a^2-4}}{2} 
\ee
This implies that $|a| \geq 2$; this could have been anticipated because the interval $I$ is mapped to $(-\infty,-2) \cup [2, \infty)$, this implies that the coefficients in the continued fraction expansion have absolute value $\geq 2$. The value $|a| =2$ leads to $\lambda = \pm 1$ which is not inside $I$, and therefore impossible. For $|a| \geq 3$ only one of the two solutions in (\ref{sol}) is in $I$; for $a$ negative the plus sign is needed, for $a$ positive one takes the minus sign.

For concreteness, let us focus on the geodesic
\bd
x = [3,3,3,\ldots] = \frac{-3 + \sqrt{5}}{2}
\ed
Chasing this geodesic leads to the following table.

\be
\ba{|r|r||c|}
\hline
\textrm{begin} & \textrm{end} & \textrm{transf.} \\
\hline
 i \infty  & \frac{-3 + \sqrt{5}}{2} & S\\
 0         & \frac{ 3 + \sqrt{5}}{2} & T^{-3}\\
 -3        & \frac{-3 + \sqrt{5}}{2} & S\\
 \frac{1}{3} & \frac{ 3 + \sqrt{5}}{2} & T^{-3}\\
-\frac{8}{3} & \frac{-3 + \sqrt{5}}{2} & S\\
   \vdots    &      \vdots & \vdots\\
\frac{3 -\sqrt{5}}{2} + \varepsilon & \frac{ 3 + \sqrt{5}}{2} & T^{-3}\\
\frac{-3 -\sqrt{5}}{2} + \varepsilon & \frac{-3 + \sqrt{5}}{2} & S\\
\vdots & \vdots & \vdots \\
\hline
\ea
\ee
Here $\varepsilon$ is a small number, which becomes smaller and smaller as we push the algorithm further and further. Though this series of begin- and endpoints never terminates, there is an easily distinguished pattern. The endpoint keeps on jumping between two values, the beginpoint converges onto another orbit of two points:
\bea
0 & \rightarrow & -3 \rightarrow [-3] \rightarrow -3 + [-3] \rightarrow [-3,-3] \rightarrow \ldots \\
& \rightarrow & [-3,-3,-3,\ldots] \rightarrow -3 + [-3,-3,-3,\ldots] \rightarrow \ldots
\eea
The corresponding geodesic segments are therefore better and better approximations of two specific geodesic segments; the consequence is that the images of the segments of the geodesic $x= \frac{-3+ \sqrt{5}}{2}$ will eventually converge on a periodic orbit. To understand this we recall two relevant facts: The natural metric on the moduli space, the Poincar\'e metric (\ref{Poin}), contains a conformal factor $1/y$, as a consequence of which the distance from every point $x+iy$ with $y > 0$ to the point $x$ is infinitely large; and, for irrational $x$ this geodesic does not land in a cusp, implying that it takes infinitely many steps to reach $y=0$ (and therefore, recalling Zeno's paradox, one may never get there).

A similar analysis as for $x=[3,3,3,\ldots]$ applies to $x=[a,a,a,\ldots]$. It is easy to see that the original limit $y \rightarrow 0$ will result in a periodic orbit in the fundamental domain. 

The analysis for sequences that are periodic with periodicity longer than 1, or become periodic after a while, is computationally slightly more involved, but conceptually the same. To find the number which represents the sequence, one uses the periodicity do derive an invariance-equation. For example $\lambda= [a,b,a,b,a,b,\ldots]$ obeys
\bd
\lambda = -\frac{1}{a-\frac{1}{b+ \lambda}} \qquad a \lambda^2 + (ab-2)\lambda-b =0
\ed
A little thought reveals that any continued fraction that ends in an infinite periodic sequence has to obey a quadratic equation \cite{Peitgen:1992}. The next check is whether any of the roots of this equation are inside the interval $I$. If so, the continued fraction is a true representative for an axion-value inside the fundamental domain. The fact that the continued fraction ends in an infinite periodic sequence implies that the mapping of the geodesic into the fundamental domain converges on a periodic orbit.

Like the rational numbers, the irrational numbers having a continued fraction expansion ending in an infinite periodic sequence are countable: They consist of a finite sequence of numbers, followed by an infinite repetition of a finite sequence. Such numbers are, like the rationals, dense in the interval $I$. 

Nevertheless, it is also clear, either from the continued fraction expansion, or from the fact that all these irrational numbers obey quadratic equations, that periodic sequences far from exhaust the irrational numbers in $I$. The ``generic'' (irrational) numbers in $I$ will give rise to an infinite, but a-periodic continued fraction expansion. Geodesics with such initial $x$ values will neither run off back to infinity, nor converge on a periodic orbit. Instead, their orbits are completely aperiodic, and never settle down. This is the chaotic realm.

High sensitivity to the initial conditions, and dense periodic orbits, are among the characteristics of chaotic behaviour \cite{Peitgen:1992}.  A third characteristic is mixing; the reader can easily convince himself of this phenomenon by computing the dual axion $a/c$ for all initial axion value $-d/c$ with sufficiently large, fixed c. That chaos takes place in this system is easy to understand from the algorithm of chasing down the geodesic; this is essentially a variant of the classic example of chaos, the stretch-and-fold transformation \cite{Peitgen:1992}. The $S$-transformation ``stretches'' (maps a finite interval to open intervals stretching to infinity), whereas the $T$-transformation ``folds'' (brings all points back to the fundamental domain.

\section{Heterotic strings on circles}\label{OPi171}

We will now turn to our original question of interest: What happens for the heterotic string on a circle when the circle size is taken to zero? 

We will set up a similar analysis as we did for $SL(2,\Z)$ duality, so first of all we need detailed knowledge of the moduli-space, which we obtain for both the $E_8 \times E_8$ and $Spin(32)/\Z_2$ descriptions. There are also a few minor discrepancies with earlier literature on the subject.

\subsection{The momentum lattice}

Both the compactifications of the heterotic string with $E_8 \times E_8$- symmetry as well as the heterotic string with $Spin(32)/\Z_2$ symmetry can be elegantly described by the formalism developed in \cite{Narain:1985jj, Narain:1986am}. 

It was realized in \cite{Narain:1985jj} that modular invariance of toroidally compactified heterotic string theory requires that the momenta of the worldsheet fields take values on an even self-dual lattice. Subsequently, in \cite{Narain:1986am} the relation between the moduli of the theory, and this self-dual lattice was clarified.

Here only compactification on a circle will be studied. The moduli are then given by the radius $R$ of the circle, and a constant gauge field $\mathbf{A}$ leading to a Wilson line. We set $\alpha'=2$, so $R$ is measured in units of $\sqrt{2/\alpha'}$; the gauge field $\mathbf{A}$ is represented as a constant vector on $\R^{16}$. In these conventions, the momenta are given by \cite{Narain:1986am}(we use linear combinations of the spatial left and right moving momenta to shorten formula's. The components given here are actually frame components.): 
\bea
\mathbf{k}_q & = & \mathbf{q} - w\mathbf{A} \\
k_n & = & \sqrt{2}\left(\frac{n+\mathbf{q} \cdot \mathbf{A} -\frac{w}{2} \mathbf{A}^2}{R}\right) \\
k_w & = & \frac{wR}{\sqrt{2}}
\eea
In these formula's $\mathbf{q}$ is a charge vector taking values on the (even, self-dual) weight lattice of the group ($E_8 \times E_8$ or $Spin(32)/\Z_2$, $n$ is the quantum number corresponding to the discrete kinematical momentum along the spatial circle, and $w$ is the string winding number. Yang-Mills theory can only be a good approximation to string theory if states with non-zero winding number have large masses; even then, some (in particular topological) aspects of heterotic string theory are starkly different from $E_8 \times E_8$ or $Spin(32)/\Z_2$ gauge theory \cite{deBoer:2001px}.
 
In these conventions the norm of the lattice is given by:
\be
\mathbf{k}_q^2 + 2 k_n k_w = \mathbf{q}^2 + 2 nw
\ee
As $\mathbf{q}^2$ takes values on an even lattice, also the lattice spanned by $(\mathbf{q};n,w)$ is even.

Expressed in the momenta, and excitation numbers $N$ and $\tilde{N}$, the mass formula for the corresponding states is:
\be \label{hetmass}
M^2 = \hlf \left(\mathbf{k}_q^2 + k_n^2 + k_w^2\right) + N + \tilde{N} - 1
\ee
The level matching constraint ensures that the mass is positive definite:
\be \label{levelm}
\frac{\mathbf{q}^2}{2} + nw + N - \tilde{N} -1 = 0
\ee

\subsection{Local and global aspects of the moduli space} \label{Ogen}

The formula's for the momenta determine the local structure of the moduli space. Taking $\R^{18}$ with an inner product given by
\be
\eta = \left(\ba{ccc} 
\mathbf{1} & 0 & 0 \\
0          & 0 & 1 \\
0          & 1 & 0 
\ea \right)
\ee
the norm on the lattice is defined by $Q^T \eta Q$, $Q^T$ being the row vector $(\mathbf{q}^T , n, w)$. The level matching constraint (\ref{levelm}) involves this $SO(17,1)$ norm. The massformula (\ref{hetmass}) is invariant under $SO(18)$-transformations, but as the level matching constraint cannot be ignored only the intersection of $SO(18)$ and $SO(17,1)$, which is $SO(17)$ can represent symmetries of the theory. $SO(17,1)$-rotations guarantee that level matching is satisfied, but change the mass-formula, and therefore generate deformations of the theory.
   
The momenta can be written as:
\be
\left( \ba{c} \mathbf{k_q} \\ k_n \\ k_w \ea \right)
=\left( \ba{ccc} 
\mathbf{1} & 0 & 0 \\
0 & \sqrt{2}/R & 0 \\
0 & 0   & R/\sqrt{2} \ea \right)
\left( \ba{ccc}
\mathbf{1} & 0 & -\mathbf{A} \\
\mathbf{A}^T & 1 & - \mathbf{A}^2/2 \\
0 & 0 & 1 \ea \right)
\left( \ba{c} \mathbf{q} \\ n \\ w \ea \right)
\ee
This is a coset representative of $SO(17,1)/SO(17)$ in the Borel gauge, and the elements of the parametrization by the Iwasawa-decomposition are explicit (see \cite{Helgason} for background on this construction).

With these assertions the local structure of the moduli-space is fixed. To determine the global structure, we have to define the group $O(\Pi_{17,1})$. It is generated by discrete transformations. In the above formula, they have to be inserted to the right of the matrices, with the discrete transformation acting on the charges, and the inverse transformation multiplying the already present matrices on the right. In general, multiplying the matrices on the right by an element of $O(\Pi_{17,1})$ may take us out of the Borel gauge. In that case, we have to multiply by an element of $SO(17)$ on the left, to return to Borel gauge. The Iwasawa decomposition theorem (see for example \cite{Helgason}) guarantees that the required element of $SO(17)$ exists.

The transformations generating the discrete group can be divided in roughly three classes. Each of these transformations leaves the spectrum of the theory invariant, but not the individual states. By relabelling the charges one can establish a formulation that leaves the mass formula and level matching formula invariant.

From the gauge theory perspective perhaps the most trivial, but more involved in heterotic string theory are the transformations we will collectively call ``shifts''. The ``shift'' is referring to the gauge field, which can be shifted by lattice vectors. In gauge theory, this is entirely trivial, as it just corresponds to the periodicity in the compact group. In the group theory language this is governed by the co-root lattice, but when dealing with simply laced groups, co-root and root lattice can be identified. For the $Spin(32)/\Z_2$, the $\Z_2$-identification implies the periodicity is affected, and elements of the coweight lattice (which, due to simply-lacedness of $Spin(32)$, can be identified with the weight lattice) enter into the periodicity considerations. We will come back to this issue when describing the moduli-space in detail.

The non-trivial aspect of string theory is that the induced transformation on the charge lattice is corrected for states with non-zero $w$. The action of the shift symmetries which we will denote by $S_{\mathbf{q}'}$ on the moduli, and the charge lattice are given by 
\be \label{shiftsym}
S_{\mathbf{q}'}: \quad
\ba{rcl}
\left(\mathbf{q};n,w\right) & \rightarrow & \left(\mathbf{q}-w\mathbf{q}', n + \mathbf{q} \cdot \mathbf{q}' - \frac{w}{2}\mathbf{q}'^ 2,w \right) \\
\left(R,\mathbf{A} \right) & \rightarrow & \left(R, \mathbf{A} - \mathbf{q}'\right)
\ea
\ee
As $\mathbf{q} \cdot \mathbf{q}'$ is integer (inner product between two lattice vectors), and $\mathbf{q}'^2$ is even, $n$ is mapped to an integer. This transformation leaves the components of the momenta separately invariant. 

For gauge theories on a circle, after gauge fixing such that the gauge fields take constant values, there is still a residual gauge symmetry due to the action of the Weyl group. The same action is present for the heterotic string on a circle, in a sense because the Yang-Mills gauge group is a subgroup of the true symmetry group of the theory. The action of a Weyl reflection $W$ is rather simple:
\be \label{Weylsym}
W: \quad
\ba{rcl}
\left(\mathbf{q}; n,w \right) & \rightarrow & \left(W(\mathbf{q});n,w \right) \\
 \left(R,\mathbf{A} \right) & \rightarrow & \left(R, W(\mathbf{A}) \right)
\ea
\ee
As $W$ is an element of $O(16)$, and the inner product is invariant under this group, invariance of the mass formula and level matching formulae follows. We will occasionally use the notation $W_{\mathbf{q}'}$ for the Weyl-reflection generated by $\mathbf{q}'$. It is worthwhile to note for future use that both for $E_8 \times E_8$ as well as for $Spin(32)$ there exists an element $W_{-1}$ of the Weyl group taking $(\mathbf{q};n,w) \rightarrow (-\mathbf{q};n,w)$, $(R, \mathbf{A}) \rightarrow (R, -\mathbf{A})$.

The Weyl group is different for the different versions of the heterotic string, $E_8 \times E_8$ or $Spin(32)/\Z_2$. Moreover, for the heterotic $E_8 \times E_8$-string, the Weyl group should be augmented with the discrete symmetry interchanging the 2 $E_8$'s, as the actual symmetry of the theory is $E_8 \times E_8 \ltimes \Z_2$ \cite{deBoer:2001px}. This can be argued on the fact that this is an exact symmetry of the theory, and string theory is not believed to have global unbroken symmetries, and therefore it must be a local symmetry. A more explicit argument is that under duality, this symmetry is mapped to one represented by an element of the Weyl group of $Spin(32)/\Z_2$, so in this description it is certainly local. Also this symmetry is an element of $O(16)$ and therefore the remarks applying to the Weyl group apply here too.

The two previous classes of transformations do not take us out of the Borel gauge, they are among the residual symmetries that this gauge allows. This is not so for the next transformation, which we will dub ``T-duality''. This is a transformation that does take us out of the Borel gauge. It may seem now that we need the explicit compensating transformation to compute the transformed moduli, but it is actually much easier to inspect the level matching and mass formula's, which are by construction invariant.

The action on the charge lattice is simply the exchange of $n$ and $w$. The level matching formula is trivially invariant under this, but the momenta obviously are not. There is not even a simple $k_n \leftrightarrow k_w$ exchange, as for example for bosonic string theory. Instead all of the momenta are mixed among each other. From the mass formula one reads off:
\be \label{Tsym}
T: \quad
\ba{rcl}
\left(\mathbf{q};n,w \right) & \rightarrow & \left(\mathbf{q};w,n \right) \\
\left(R,\mathbf{A} \right) & \rightarrow & \left(\frac{2R}{R^2+\mathbf{A}^2}, -\frac{2\mathbf{A}}{R^2+\mathbf{A}^2}\right)
\ea \ee
The transformation $TW_{-1}$ has a fixed surface at $R^2 + \mathbf{A}^2 = 2$.
 
A last, rather trivial discrete symmetry is parity 
\be \label{parsym}
P: \quad
\ba{rcl}
(\mathbf{q},n,w) & \rightarrow & (\mathbf{q},-n,-w) \\
(R,\mathbf{A})& \rightarrow & (R,-\mathbf{A}).
\ea \ee
Parity on the compactification circle inverts both momenta and winding. It acts on $\mathbf{A}$, because under parity the holonomy around the circle is exchanged for its inverse. The parity action can be combined with the symmetry $W_{-1}$ to obtain a symmetry acting as inversion on the charge lattice, leaving the moduli invariant. This symmetry is analogous to the element $-\mathbf{1}$ in $SL(2,\Z)$, that rotates the charge lattice over an angle of $\pi$, but has no effect on the scalar moduli. 

\subsection{Explicit description of the moduli-spaces of heterotic strings}

From the previous section, the moduli space arises in the form
\be
SO(17) \backslash O(17,1) / O(\Pi_{17,1})
\ee
We have given different descriptions of $O(\Pi_{17,1})$ for $E_8 \times E_8$ and $Spin(32)/Z_2$; The subgroups corresponding to shifts and Weyl-reflections depend on the geometry of the root lattices of the corresponding groups. It is a standard result (that we will rederive in section \ref{SpinE8dual}) that the resulting moduli-spaces are nevertheless isomorphic. We will now give a detailed description of the moduli-space in ``$Spin(32)/\Z_2$- coordinates'' and ``$E_8 \times E_8$-coordinates''. Though formulated slightly more abstractly, the ``$Spin(32)$-version'' is essentially identical to the one appearing in \cite{Cachazo:2000ey} (up to a minor difference that we will mention later).

We will start by concentrating on those discrete symmetries that can be associated to the presence of the gauge group. For Yang-Mills theories on a circle (of fixed radius), the moduli consist of the holonomy of the gauge field $\mathbf{A}$ around the circle. This holonomy is an element of the gauge group, and only defined up to conjugation by elements of that same group. This freedom to conjugate with group elements allows to ``diagonalize'' the holonomy, that is to transform it into an element of the maximal torus (the abelian subgroup obtained by exponentiating the Cartan subalgebra of the gauge group). These elements can be parametrized by elements of the Cartan subalgebra, subject to a periodicity condition (due to the compactness of the maximal torus). There is a residual symmetry, the Weyl group of the gauge group normalizes the maximal torus, and hence the fundamental domain for the torus. Consequently, the moduli space is an orbifold of the maximal torus by the Weyl group. The detailed structure depends of course on the gauge group. In case the gauge group obeys a number of conditions, a neat description is given by the following theorem, taken from \cite{Kac:1999gw}\footnote{The rough contents of this theorem have been used earlier and in other places. The formulation of \cite{Kac:1999gw} is precise and suits our needs.}.

We first fix some definitions. Let $G$ be a connected simply connected (almost) simple compact Lie group of rank $r$. To the algebra of this group there is associated a basis of simple roots $\alpha_i$ ($1 \leq i \leq r$). These simple roots can be graphically encoded in a Dynkin diagram, and span the root lattice. The height of a root is given by the sum of its expansion coefficients, when expanded on the basis of simple roots. There is a unique highest root $\alpha_H$, which has the expansion (where actually $a_0$ is always 1)
\be
a_0 \alpha_H = \sum_{i=1}^r a_i \alpha_i
\ee
The dual lattice to the root lattice is spanned by the fundamental coweights $\omega_i$, defined by $\langle \alpha_i, \omega_j \rangle = \delta_{ij}$ (with $\langle \cdot, \cdot \rangle$ the natural bilinear form on the root space. 

We have the following 

{\bf Theorem}: For a set of $r + 1$ real numbers $\vec{s} = (s_0, s_1,\ldots , s_r)$ such that 
\be
s_j \geq 0, \qquad j = 0,\ldots, r; \qquad \sum_{j=0}^r a_j s_j =1
\ee
consider the element
\be
\sigma_{\vec{s}} = \exp \left(2 \pi i \sum_{j=1}^r s_j \omega_j \right)
\ee
one has
\begin{enumerate}
\item Any element of $G$ can be conjugated to a unique element $\sigma_{\vec{s}}$.
\item The centralizer of $\sigma_{\vec{s}}$ (in $G$) is a connected compact Lie group which is a product
of $(U(1))^{n-1}$ where $n$ is the number of non-zero $s_i$, and a connected semi-simple group whose Dynkin diagram is obtained from the extended Dynkin diagram  for $G$ by removing the nodes $i$ for which $s_i\neq 0$.
\end{enumerate}

There is a third part to the theorem in \cite{Kac:1999gw} that we will not need here (it deals with the topology of the unbroken gauge group, and the version of \cite{Kac:1999gw} is not valid in string theory due to the presence of winding states. For details see \cite{deBoer:2001px}). If the heterotic gauge groups obeyed the conditions of the theorem, then we would have a parametrization of the moduli space by the parameters $s_1,\ldots s_r$.  The conditions of the theorem are however violated, but this can be repaired by a little more analysis. We will first treat $E_8 \times E_8$ and then $Spin(32)/\Z_2$

The group $E_8 \times E_8$ is not simple, but this imposes no difficulty; the moduli space for $E_8 \times E_8$-theory on a circle is simply given by the direct product of 2 copies of the moduli-space for $E_8$ Yang-Mills theory on a circle. More severe is that the actual gauge group is actually $E_8 \times E_8 \ltimes \Z_2$, where the $\Z_2$ acts as an interchange of the 2 $E_8$-factors. As a consequence, this group is not connected. Nevertheless, it is not to hard to account for this. 

Holonomies in the disconnected component of $E_8 \times E_8 \ltimes \Z_2$ lead to a construction now commonly known as ``CHL-strings'' \cite{Chaudhuri:1995bf}\footnote{The abbreviation ``CHL'' refers to the authors of \cite{Chaudhuri:1995fk}. In \cite{Chaudhuri:1995bf} the construction as we formulate it was presented. As an orbifold, versions of the CHL-string appear to have been invented more than once, see for example the introductory section of \cite{Schellekens:1989dk} for an early version.}. We only consider the connected component here.
With the above parametrization, we introduce parameters $t_1$ to $t_9$, and $t_{11}$ to $t_{19}$. The choice of indices is suggested by making a correspondence to the nodes of the Dynkin-diagram of $\Pi_{17,11}$, and its $E_8 \times E_8$-basis, as in appendix \ref{latcon}. The gauge theory description should be valid for large $R$; consequently we delete node 10 (which corresponds to the enhanced symmetry appearing at the critical radius) and are left with two extended $E_8$ diagrams.  Transcribing the above theorem for the situation of $E_8 \times E_8$ (lattice conventions for $E_8$ can be found in appendix \ref{appe8}), the moduli space is parametrized by
\be
t_i  \geq 0 
\ee
\be
2t_1 + 4t_2 + 3 t_3 + 6 t_4 + 5 t_5 + 4 t_6 + 3 t_7 + 2 t_8 +t_9 =1
\ee 
\be
2t_{19} + 4t_{18} + 3 t_{17} + 6 t_{16} + 5 t_{15} + 4 t_{14} + 3 t_{13} + 2 t_{12} + t_{11} =1
\ee 
The two latter relations allow us to eliminate 2 of the 18 parameters, so that there are in total 16 free parameters. 

The extra $\Z_2$ acts as an interchange of the two $E_8$-factors. Consequently, we can demand in addition
\be
t_9 \leq t_{11}
\ee
This condition does not eliminate any parameters.

The explicit Wilson line, with the $t_i$ constrained as above, is (we use the superscript $R$ in this article to denote a vector, with the order of its components reversed)
\be
A = -\sum_{i=1}^{8} t_i \omega_{9-i}^R + \sum_{t=12}^{19} t_i \omega_{i-11}
\ee

According to the theorem, the moduli space has singularities whenever any of the $t_i$ is $0$. At such points there is enhanced symmetry, given by deleting from the Dynkin diagram the nodes $i$ for which $i \neq 0$. These singular loci are the fixed planes for Weyl reflections (if $t_9$ or $t_{11}$ is involved, a combination of Weyl reflection and shift), and are hyperplanes cutting up the covering space in fundamental domains. 

Another boundary of the moduli space is given by the equation $t_9 = t_{11}$. This is more like the boundary of the fundamental domain $SL(2,\Z)$ moduli space. It does not represent a singular surface in itself, but passes through the 8-dimensional singular locus 
\be
(t_1, t_2, t_3, t_4, t_5, t_6, t_7, t_8) = (t_{19}, t_{18}, t_{17} t_{16}, t_{15}, t_{14}, t_{13}, t_{12})
\ee  
which implies $t_9 = t_{11}$. At this 8-dimensional locus there is an enhanced discrete symmetry (and being discrete, there are no massless gauge bosons associated to it), to be precise a $\Z_2$ symmetry. It is at this precise locus that for example the CHL-construction \cite{Chaudhuri:1995bf} is possible. It is crucial for the T-duality between the two kinds of heterotic string, that a similar singularity occurs in the $Spin(32)/\Z_2$-description. We will now show that this is indeed the case.

The group $Spin(32)/\Z_2$ is simple and connected, but it is not simply-connected, and therefore also needs additional analysis. We introduce 17 parameters $t_2', \ldots t_{18}'$. Their possible values are determined by 
\be
t'_i \geq 0
\ee
\be \label{Spinconstr}
t'_2 + t'_3 + t'_{17} + t'_{18} + 2\sum_{i=4}^{16} t'_i =1
\ee
where again the relation allows to eliminate 1 parameter, so that there are again 16 parameters. This would be for $Spin(32)$-theory, but we are dealing with a $Spin(32)/\Z_2$ theory. This implies an extra identification under the shifts $\mathbf{A} \sim \mathbf{A} + \omega_{16}$. Combining this action with Weyl-group elements leads to 
\be \label{spinz2action}
\mathbf{A} \rightarrow \omega_{16} -\mathbf{A}^R.
\ee 
This maps the fundamental domain to itself \cite{Schweigert:1996tg} (an explicit computation demonstrating this can be found in \cite{Keurentjes:2000mk} section 4.3). The action induced on the $t'_i$'s is:
\be
t'_i \rightarrow t'_{20-i}
\ee
We may therefore choose our gauge field to be on one side of the hyperplane halfway between the null-vector and $\omega_{16}$. The equation of this hyperplane is given by those $\mathbf{v}$ obeying
\be
\mathbf{v} \cdot \omega_{16} \leq 2
\ee 
Expressing $\omega_{16}$ in the roots, and substituting $\mathbf{v} = \sum_{i=3}^{18} t'_i \omega_{i-2}'$, this equation becomes
\be
\sum_{n=3}^{16}(n-2) t'_n + 7 t'_{17} + 8 t'_{18} \leq 4
\ee
Subtracting 4 times the constraint equation (\ref{Spinconstr}) gives the following attrative form of the equation
\bd
4 t'_2 + 3t'_3 + 6 t'_4 + 5 t'_5 + 4 t'_6 + 3 t'_7 + 2 t'_8 + t'_9 \geq
\ed
\be t'_{11} + 2 t'_{12} + 3 t'_{13} + 4 t'_{14} + 5 t'_{15} + 6 t'_{16} + 3 t'_{17} + 4 t'_{18}
\ee
in which we recognize the coefficients for the linear dependence relation for $\Pi_{17,1}$ again. To make them appear, we had to eliminate $t'_{10}$ from the equation; node number 10 does not appear in the linear dependence relation. 

Also now, not all of this hyperplane represents a singularity. There is an honest singularity for
\be
(t_2, t_3, t_4, t_5, t_6, t_7, t_8, t_9) = (t_{18}, t_{17}, t_{16}, t_{15}, t_{14}, t_{13}, t_{12}, t_{11})
\ee
which is again an 8 dimensional locus, with enhanced, discrete $\Z_2$ symmetry\footnote{Puzzled or inexperienced readers may compare with $SO(3)=SU(2)/\Z_2$. The centralizer of a generic element of $SO(3)$ is $SO(2)$, but elements conjugate to $diag(-1,-1,1)$ have centralizer $O(2)$; also here there is symmetry enhancement by a discrete $\Z_2$}.

With the constraints taken into account, the $t'_i$ parametrise the Wilson line
\be
\mathbf{A} = \sum_{i=3}^{18} t'_i \omega'_{i-2}
\ee 

\FIGURE{
\includegraphics[width=6cm]{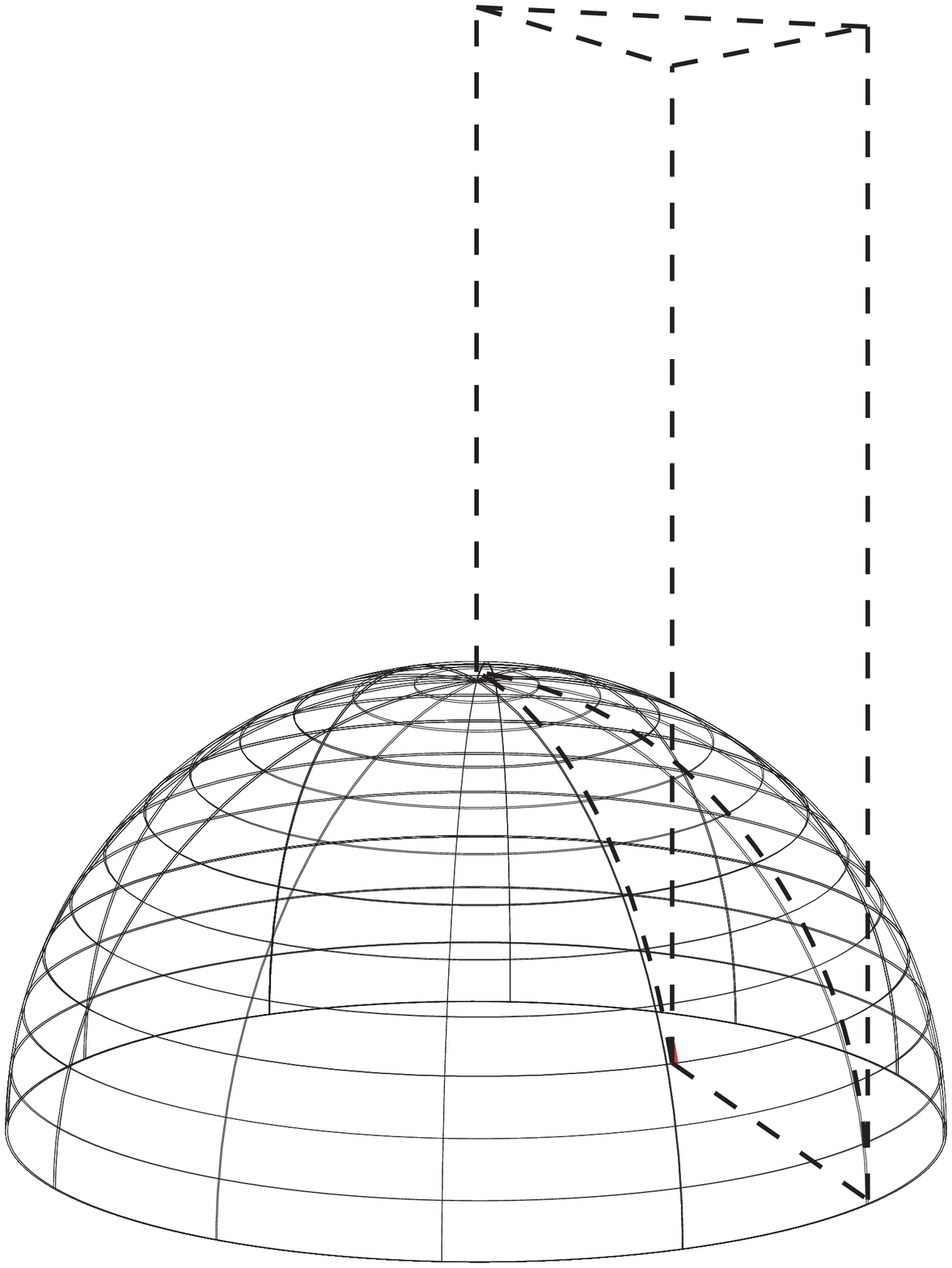}
\caption{Impression of the moduli space.}\label{modspace}
}
In both the $E_8 \times E_8$ and $Spin(32)/\Z_2$ cases, after having taken the shifts, Weyl-reflections and other discrete gauge symmetries into account, we are dealing with a moduli space which is schematically of the form ``convex polygon divided by $\Z_2$-symmetry times $\R$'' (this is the ``chimney'' in the words of \cite{Cachazo:2000ey}). The symmetry not yet taken into account is T-duality. The surface fixed under T-duality (actually, in our conventions $TW_{-1}$) is given by the equation 
\be
R^2 + \mathbf{A}^2 = 2
\ee
which will form a lower boundary of the fundamental domain. 

Of course it is now important to check whether the``chimney'' fits into a circle of radius 2. This may seem tedious to check, but is actually not so difficult; because of the convexity of the polygon, the distance function $\mathbf{A}^2$ will take its maxima at the corners of the polygon, of which there are only finitely many. The result in both the $E_8 \times E_8$ description and the $Spin(32)/\Z_2$ description is the same: The whole polygon has 
\be
\mathbf{A}^2 \leq 2
\ee
with equality for a single (corner-)point only. This point is given by $\omega_E$ for the $E_8 \times E_8$ description, and by $\omega_S$ for the $Spin(32)/\Z_2$ description. 

Figure \ref{modspace} is an impression of the moduli space, with 14 dimensions suppressed. The ``base'' of the chimney rests on a single ball. That this fundamental domain contains only a single point at $R=0$ is important for understanding the duality between heterotic $Spin(32)-E_8 \times E_8$ T-duality, but also for the chaotic properties to be analyzed in the next section. If the single point at $R=0$ were not a single point, but a surface, it would be much ``easier'' to hit than the decompactification limit $R \rightarrow \infty$; but then the duality could not possible be symmetric, and the dual theory would have an entirely different character (it could not be the dual of something that is approximately a gauge theory on a circle of large radius). Note also that this would imply that the moduli-space has an infinite volume, and would violate the ``swampland'' conjectures of \cite{Vafa:2005ui}. On the other hand if even this single point were absent, there would only be a single description, and never a duality to a theory with a different gauge group. 

\subsection{The ``fiducial'' moduli-space of type I strings on a circle}

With the present technology, it is also easy study the points made in  \cite{Cachazo:2000ey} on the reach of perturbation theory in type I and type I'(starting from some puzzles raised in \cite{Bachas:1997kn}). First however we should explain the difference between our figure \ref{modspace} and figure 2 of \cite{Cachazo:2000ey}.
 
The description in \cite{Cachazo:2000ey} is almost identical to ours (though phrased differently), but the authors do not divide by the $\Z_2$-action (\ref{spinz2action}), which is called an outer automorphism there. We do insist on dividing by this $\Z_2$ because in fact, in the $Spin(32)/\Z_2$ description the relevant automorphism is \emph{inner}. We note that $Spin(32)/\Z_2$ does not have an outer automorphism, as the outer automorphism of $Spin(32)$ does not descend to $Spin(32)/\Z_2$, as for example explained in \cite{deBoer:2001px}). The action (\ref{spinz2action}) can be composed out of Weyl reflections generated by $e_i \pm e_j$  (these generate permutations of the components of the Wilson, as well as multiplication of components by minus signs) with a shift; both Weyl reflections and shifts correspond to inner automorphisms, and hence their composition does too.

There are ony discrete gauge symmetries associated to this extra symmetry, and therefore no extra massless particles. Hence, only some details change, as the description of \cite{Cachazo:2000ey} is effectively a double cover of the fundamental domain. 

In \cite{Cachazo:2000ey} it is observed that type I (and type I') perturbation theory does not reach into all corners of the moduli space. Though at first sight, taking dualities into account, this may seem puzzling, there is a straightforward explanation for this (though our version does not exactly coincide with the one given in \cite{Cachazo:2000ey}). The gauge group that is manifest in type I perturbation theory is $O(32)/\Z_2$ (The $\Z_2$ is generated by the element $\mathbf{-1}$ of $O(32)$) instead of $Spin(32)/\Z_2$. First of all, this is a disconnected group, and naive consideration of the moduli space would lead to two irreducible components. We will only consider the component connected to the identity, which parametrizes elements of $SO(32)/\Z_2 = Spin(32)/(\Z_2 \times \Z_2)$. We will call the space of $SO(32)/\Z_2$ flat connections the ``fiducial'' moduli-space of type I.

The fiducial moduli space of type I has roughly the same structure as the one considered for $Spin(32)/\Z_2$, but we have to take an extra $\Z_2$ identification into account. The particular $\Z_2$ is generated by shifts over $\omega_1$ (it is easily verified that integer shifts over $\omega_1$ have no effect on states in the adjoint and symmetric tensor representations of $SO(32)$, which are the only ones appearing in type I perturbation theory). To find the action on the fundamental domain, we have to combine the shift with an appropriate element of the Weyl group. The particular element of the Weyl group (in our conventions) multiplies the first and the last component of $\mathbf{A}$ by $-1$. Denoting this action by $\theta$, the transformation acts on the Wilson line as
\be
\mathbf{A} \rightarrow \omega_1 - \theta(\mathbf{A}).
\ee
The action induced on the $t'_i$ is
\be
t'_0 \leftrightarrow t'_1, \quad t'_{15} \leftrightarrow t'_{16}
\ee

We do as before, and restrict the gauge field to be one side of a hyperplane through moduli space, given by
\be
\mathbf{v} \cdot \omega_1 \leq \hlf
\ee
To translate this to the description given earlier, we have to substitute $\mathbf{v} = \sum_{t=1}^{16} t'_i \omega_i$, and expand $\omega_1$ in the simple roots. The result gives
\be
2 \sum_{i=1}^{14} t_i + t_{15} + t_{16} \leq 1
\ee   
or, taking the constraint into account
\be \label{Icond}
t_1 \leq t_0
\ee

Qualitatively, the fiducial moduli space for type I is similar to the heterotic moduli-space, except for the fact that the ``chimney'' is smaller. The hypersurface $t_1 = t_0$ passes through the subspace with equation
\be
(t_1,t_{15}) = (t_0,t_{16})
\ee
which is a singularity of the fiducial moduli-space. At the singularity there is a discrete $\Z_2$ symmetry acting among the perturbative states of type I. 

The punchline is of course that this singularity does \emph{not} occur in the ``real'' type I moduli space, because the $\Z_2$ symmetry does not extend to the non-perturbative states. From heterotic-Type I duality, it is known that there must be states in one of the spinor-representations of $Spin(32)$ \cite{Sen:1998tt, Witten:1998cd}; the above $\Z_2$ action would map these to states of the \emph{other} spinor representation, which is not present.

At a cross-section of the chimney, for fixed sufficiently large $R$, the slices of the moduli-space are a double cover of the fiducial moduli space. Below $R^2 \leq 2$ things get more involved, as the $\Z_2$ action that defines the fiducial moduli-space does not preserve $\mathbf{A}^2$.  This leads to the ``missing'' regions of moduli-space of \cite{Cachazo:2000ey}, where perturbation theory is supposed to have broken down. However for perturbation theory, defined as the systematic way to compute correlators where no non-perturbative states are allowed as initial, intermediate or final state, the ``fiducial'' moduli space is all one needs, and there are no mysteries there. In particular there are no ``missing'' regions; according to the identifications that make up the fiducial moduli space these are mapped back into the fundamental domain. 

Of course using the fiducial moduli-space is fundamentally wrong, the $\Z_2$-symmetry (\ref{spinz2action}) is broken by non-perturbative states, and including these must inevitably lead to discarding the fiducial moduli space, and the embracing of the ``real'' moduli-space.

For heterotic $Spin(32)$-theory at large radius, the fiducial moduli-space gives an acceptable approximation for many computations, because the states breaking the $\Z_2$ symmetry have large masses (the lightest are of order $(\alpha')^{-\hlf}$). The region that is deemed ``missing'' in \cite{Cachazo:2000ey} is the region where states in spinor representations of $Spin(32)$ acquire masses of the same order as the ``light'' states in the spectrum, and hence it is inconsistent to leave them out. 
 
\section{The zero-radius limit}

Having gathered all the necesary information on the moduli-space, it is now time to take the $R \rightarrow 0$ limit.

\subsection{A slice of the moduli space}

In this section we concentrate on a \mbox{2-di\-mensio\-nal} slice of the moduli space, parametrized by the radius of the theory and special Wilson lines. It is in many aspects representative for the whole fundamental domain.

There are actually two cases that can be discussed almost simultaneously. The first case is heterotic $E_8 \times E_8$-theory, on a circle of radius $R$ with a Wilson line $A_E = \lambda \omega_E$. Due to the discrete identifications (shifts and Weyl reflections) that define the moduli-space, we can restrict the values of $\lambda$ to $0 \leq \lambda \leq 1$. A qualitatively and quantitatively similar picture arises when considering heterotic $Spin(32)/\Z_2$-theory, on a circle of radius $R$, with a Wilson line $\lambda \omega_S$, $0 \leq \lambda \leq 1$. For concreteness, we will focus on heterotic $E_8 \times E_8$-theory, and metion at the end of the analysis the modifications needed for $Spin(32)/\Z_2$-theory. 

The particular slice from the moduli-space we have selected runs exactly to the single point in the fundamental domain with $R=0$. It is convenient to introduce the rescaled radius $r = R /\sqrt{2}$. The analysis is similar to the $SL(2,\Z)$-situation. There are 3 relevant transformations, acting on the pair $(r,\lambda)$.

\FIGURE{
\includegraphics[width=7cm]{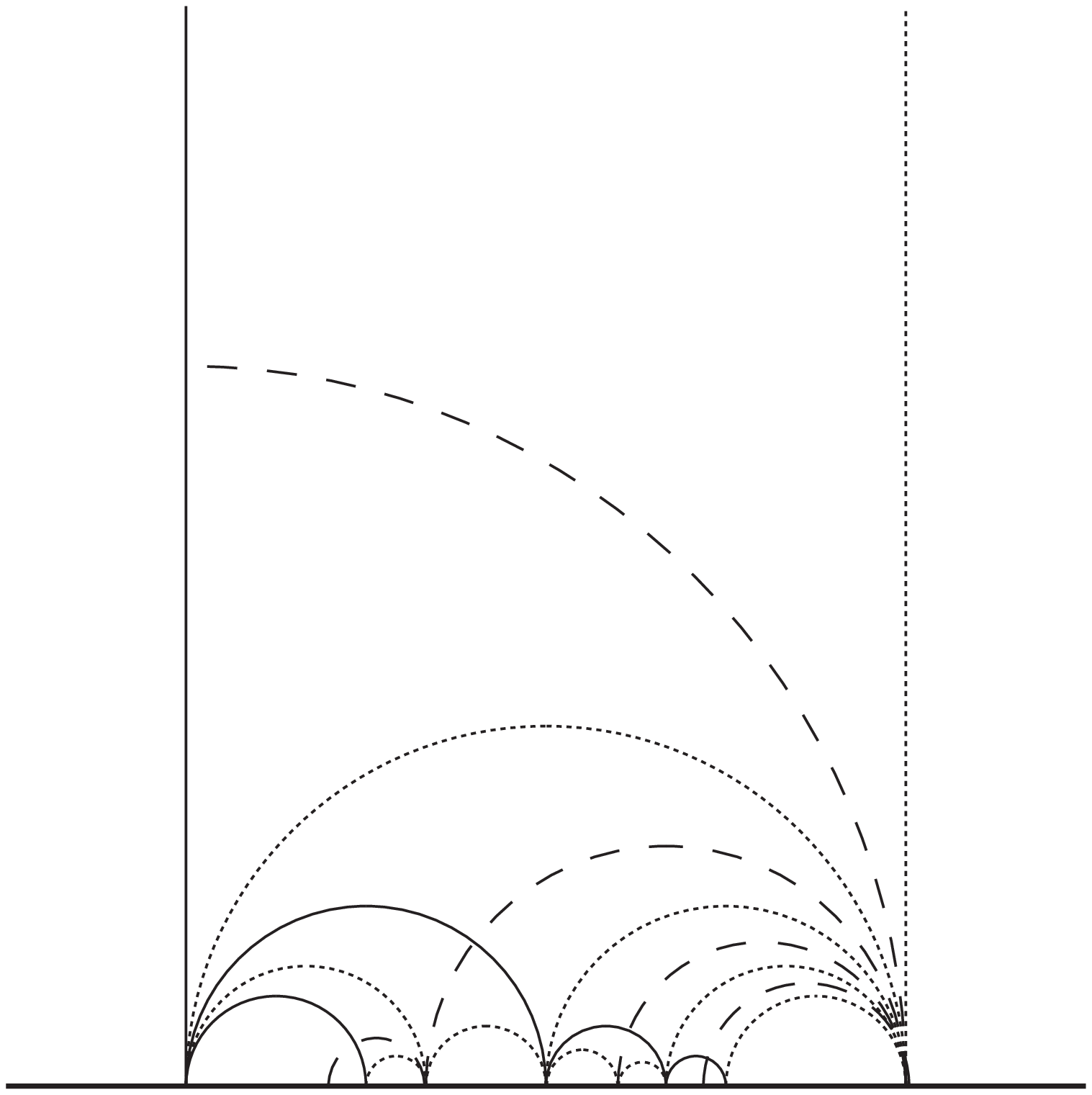}
\caption{A 2-dimensional slice of the moduli-space. See text for details.}\label{hetmod}
}

First there are the shifts: As $2 \omega_E$ is an element of the root lattice 
\bd (r,\lambda) \rightarrow (r, \lambda \pm 2) \ed
defines an identification on this slice of the moduli-space.

Second, there are the Weyl group elements. The only fact that we will need, is that there are among these elements that inverts the Wilson line, and hence send 
\bd
(r,\lambda) \rightarrow (r,-\lambda).
\ed

Third, the action of T-duality is 
\bd
(r,\lambda) \rightarrow \left(\frac{r}{r^2 + \lambda^2}, \frac{-\lambda}{r^2 + \lambda^2}\right). \ed

The fixed lines under these symmetries (and combinations of them) are drawn as solid and dashed lines in figure \ref{hetmod}. At each of these lines there is enhanced gauge symmetry. At the solid line there is $E_8 \times E_8$-symmetry, at the line with small dashes there is an unbroken $SO(16) \times SO(16)$, at the line with large dashes there is $SO(14) \times SO(14) \times SU(2)$ symmetry. In the interior, the typical unbroken group is $SO(14) \times SO(14)$. In all these cases $U(1)$'s have to be added to enhance the rank of the group to 18.

When the solid and small dashed line meet, there is enhanced $E_8 \times E_8 \times SU(2)$ symmetry. The meeting of the solid and large dashed line signals the decompactification to the $E_8 \times E_8$ theory, the meeting of small and large dashes implies decompactification to $Spin(32)/\Z_2$-theory.
 
If we are only interested in what the fate is in the limit $r \rightarrow 0$, we can simply set $r=0$ from the outset. Then we only have to worry about $\lambda$ which can be transformed to $\lambda \pm 2$, $-\lambda$, and $1/\lambda$.   

We will use an algorithm that is an adaptation of the one we used for $SL(2,\Z)$ duality. The limit $r \rightarrow 0$ at fixed Wilson line corresponds to a geodesic. Using a combination of shifts and the Weyl reflection, we can ensure that this geodesic has $0 \leq \lambda \leq 1$. When it gets too close to $r \rightarrow 0$, we apply a T-duality, sending the endpoint $(0,\lambda) \rightarrow (0,-1/\lambda)$. When $0 < \lambda < 1$, $-1/\lambda$ is outside the fundamental domain, and shifts and Weyl-reflections have to be applied to get back to $0 \leq \lambda \leq 1$.

A crucial difference with the $SL(2,\Z)$ story is that there are now \emph{two} points where this algorithm can end (if it terminates at all). One of the two is $\lambda = 0$; after a single T-duality this corresponds to a decompactification to the 10-dimensional $E_8 \times E_8$-theory. The second possibility is $\lambda =1$. This the $R\rightarrow 0$ limit of the theory with Wilson line $\omega_E$, and hence corresponds to the decompactification to the $Spin(32)/\Z_2$ theory.

We now study the pre-images of the above end-points. Under shifts and Weyl reflections, the $E_8 \times E_8$ endpoints, all correspond to $\lambda=even$, whereas the $Spin(32)$-endpoints are given by $\lambda =odd$. Including the T-duality in the set of transformations leads to the following picture. 

It is clear that the sequence of shifts, reflections and inversion will terminate if an only if $\lambda$ is rational, and hence can be written as $p/q$, with $\gcd(p,q)=1$. Because the latter constraint, we can distinguish two subclasses: $p$ and $q$ are both odd; or, either $p$ or $q$ is even, and the other odd. The shifts, reflections and inversions acts within each of these two classes separately, and therefore the case $\lambda$ rational, $p$ and $q$ both odd will end on $\lambda =1/1$, and hence on the decompactified $Spin(32)$ theory, whereas the second class can be transformed to $\lambda =0/1$, and hence will end on the decompactified $E_8 \times E_8$-theory.

Of course the rationals only represent a countable subset of the interval $0 \leq \lambda \leq 1$, and again, the generic $\lambda$ is irrational. For these it is convenient to temporally ignore the reflection $\lambda \rightarrow -\lambda$, and work on the bigger interval $-1 \leq \lambda \leq 1$. This is possible, because commuting the inversion with a shift results in another shift, and the inversion commutes with T-duality. As a consequence, one can commute all inversions out of the intermediate steps, and deal with the net number of inversions in the end.

The computation is then exactly as for the $SL(2,\Z)$ case, with the exception that the domain for $\lambda$ is twice as large as for $\textrm{Re}(\tau)$\footnote{The computation for $\lambda$ would correspond to the situation where we are not dividing by $SL(2,\Z)$, but by its subgroup $\Gamma(2)$.}. Apart from that, the same algorithm using continued fractions can be set up. Again there is the interesting subcase that such a continued fraction expansion may end in a tail with periodicity properties. Such a situation leads to the same conclusion, that the corresponding geodesic converges onto a periodic orbit in the fundamental domain. This conclusion is unaffected by the fact that in the end we still have to account for the reflections, as these cannot change the fact that the orbit is periodic.
 
It is easy to see that also now the set of geodesics converging onto periodic orbits is countable, and that therefore the majority of geodesics, when mapped to the fundamental domain, bounces about chaotically, and never converges to either a decompactification limit nor a periodic orbit.

Essentially the same discussion applies to heterotic $Spin(32)/\Z_2$-theory on a circle with $\lambda \omega_S$ (This is not the dual image of the previously discussed slice of the $E_8 \times E_8$ moduli space!), provided we modify a few things. Now the solid line corresponds to unbroken $Spin(32)$, the small dashed line corresponds to unbroken $SO(16) \times SO(16)$, and the large dashed line to $SO(14)\times SO(16) \times SU(2)$. In the interior, the symmetry is $SO(14) \times SO(16)$. When solid and small dashed lines meet, there is enhanced $SO(32) \times SU(2)$-symmetry. The meeting of solid and small dashed lines now corresponds to decompactification to the $Spin(32)/\Z_2$ theory, and the meeting of small and large dashes to decompactification to the $E_8 \times E_8$ heterotic theory. The rest of the above analysis remains the same, with the same transformations and conclusions, except that the roles of the $E_8 \times E_8$ and $Spin(32)/\Z_2$ theories are interchanged. 

\subsection{Determining the dual}

Extending the procedure of the previous section to the whole of the moduli-space is simple. Again we compute the fate of endpoints (and beginpoints of geodesics under a sequence of discrete transformations.

The different transformations are known, and consists of Weyl-reflections, shifts and the T-duality transformation. As in the previous section, it is possible to ignore the Weyl group for the time being, for the same reasons: Weyl reflections commute with T-duality, while commuting a shift through a Weyl group element results in a Weyl -transformed shift. We therefore temporarily extend the fundamental domain with all its Weyl-transformed copies. As a consequence of our description of the moduli space, this extension fits within the circle with equation
\be
\mathbf{A}^2 \leq 2
\ee
and touching the circle at finitely many points (which are the orbits of $\omega_E$ or $\omega_S$ under the $E_8 \times E_8$ and $Spin(32)$ Weyl group respectively).

We now build essentially a higher dimensional version of the algorithm we have used before. Taking the limit $R \rightarrow 0$ with fixed $\mathbf{A}$ corresponds to moving along a geodesic, that we can assume to start within the fundamental domain. The endpoint of the geodesic is outside the fundamental domain (it is inside the chimney, but also inside the ball). We then perform a T-duality, to take the end-point outside the ball, and then apply a sequence of appropriate shifts, to take the endpoint back inside the ball. If the endpoint is now inside the ball, the process repeats.

Again there are two possible endpoints for such a sequence of transformations. Either one ends up at $\mathbf{A} =0$, in which case one can by a T-duality transform to the T-dual theory at $R= \infty$ which is then automatically the same theory as before. Or after finitely many steps one ends up exactly at the sphere of radius 2. If it is impossible to shift the endpoint into the ball from there, then the inevitable conclusion is that we must be at one of the images of the (up to Weyl-transformations) unique point that will lead to decompactification to the other theory.

If the endpoint implies decompactification to the same theory, the Wilson-line in the corresponding theory can be computed by backtracking the beginpoint at infinity through the sequence of duality transformations. If the endpoint implies decompactification to the other theory, then it is still usefull to track the beginpoint; With the map between the two heterotic theories of section \ref{SpinE8dual} one can find the Wilson line in the dual theory.

The above algorithm boils down to a sequence of shifts $S_{\mathbf{q}}$ over some vector, separated by T-duality transformations $T$. It is convenient to introduce a shorthand notation similar to the one we used for $SL(2,\Z)$:
\be
[\mathbf{q}_1, \mathbf{q}_2, \mathbf{q}_3, \ldots] \qquad \leftrightarrow \qquad TS_{\mathbf{q}_1}TS_{\mathbf{q}_2}TS_{\mathbf{q}_3}T \ldots
\ee
It is then possible to distinguish between finite and infinite sequences of $\mathbf{q}_i$'s. The finite sequences again lead to one of the two decompactification limits. The infinite sequences can either have a periodic tail, or be completely a-periodic, leading to convergence to periodic orbit, or a chaotic orbit respectively.

It is obvious to see that any finite sequence corresponds to a Wilson line which, when expanded on a basis of simple coroots (or simple coweights), has rational coefficients. We suspect that the converse statement, that any Wilson line having an expansion of coroots with rational coefficients corresponds to a finite sequence, is also true, but we know no proof (for previously discussed cases, where the axion or Wilson line takes values in a 1 dimensional (sub-)space, simple proofs can be given using the ordering of points on a line. For the higher dimensional case it is not obvious how to generalize this). In this case the Wilson lines would correspond to ``rational'' points.

For irrational points with sequences with periodicity properties, an invariance equation exists. For sequences where the periodicity is one (ending in an infinite sequence of copies of the same vector) the situation is analogous to the 1 dimensional case. The general case is, by its higher dimensional nature, much harder to analyze.

\subsection{Finite radius}

Every point $p$ of the moduli-space that is outside the fundamental domain, but not at $R=0$ can be mapped into the fundamental domain by a sequence of shifts, Weyl reflections and T-dualities. One can take the straight geodesic running through both the point $p$ and $R = \infty$, and apply an appropriate transformation every time one runs into a boundary of the moduli-space. It is however important to stress that such an algorithm does \emph{not} (necessarily) coincide with the procedure we have sketched in the previous section, as our procedure to determine the $R \rightarrow 0$ limit does not necesarily result in geodesic segments that always lie within the fundamental domain.

The algorithm that produces a sequence of geodesic segments staying inside the fundamental domain will, if continued to $R = 0$, obviously have the same begin- and endpoint as our procedure. The ambiguity at the intermediate stages is caused by the algebraic relations existing among the discrete group. For $PSL(2,\Z)$ one has the relations (\ref{SLrel}), $O(\Pi_{17,1})$ has (among others) relations of the form
\be \label{Orel}
(TS_{\mathbf{q}})^3 = W_{\mathbf{q}} P \qquad \textrm{if }\mathbf{q}^2 = 2
\ee
Such relations allow to write the transformation leading from begin to endpoint in various forms, leading to different trajectories for the geodesics on the covering space.

It will however be clear that such ambiguities in no way affect the qualitative results of chaotic behaviour: For sufficiently small $R$, the map to the fundamental domain for the pair $(R,\mathbf{A})$ gets extremely sensitive to small variations in $\mathbf{A}$ and $R$.

As an aside, the relation (\ref{Orel}) implies that $O(\Pi_{17,1})$ can be generated from a set of elements that need not contain Weyl group generators. This is amusing to see, as it is the Weyl reflections that received a very prominent place in our analysis (determining the shape of the chimney); but as long as there is the T-duality transformation, they can be regarded as secondary.

\subsection{From $E_8 \times E_8$ to $Spin(32)/\Z_2$ and back}\label{SpinE8dual}

An explicit map relating the charge lattices for the $Spin(32)/\Z_2$ formulation and the $E_8 \times E_8$ version was already given in \cite{Ginsparg:1986bx}. The map between the moduli seems not to have been written down for the general case, although there are many studies of special cases.

Previous formulations of the duality map seem to require some educated guesses. When explicit bases for the $\Pi_{17,1}$ in $Spin(32)$- and $E_8 \times E_8$-coordinates are given, it is not hard to derive the exact map explicitly.

We have given our conventions for the two coordinate-systems describing the same lattice in appendix \ref{latcon}.

Consider now a state, in the $Spin(32)/\Z_2$-formulation, which has charges $(\mathbf{q},n,w) = (\mathbf{0},1,0)$. Level matching then implies $N-\tilde{N} =1$. Though the following computation can be done for any value of $N$, we will set $N=1$ ($\tilde{N} = 0$). The charges $(\mathbf{q};n,w) = (\mathbf{0},1,0)$ imply, in the conventions of appendix \ref{latcon}, that this state corresponds on the $E_8 \times E_8$-side, to a set of charges $(-2 \omega_E;2,-2)$. We can then compute the masses in both pictures (because ground states of the oscillators on one side of the duality have to correspond to ground states on the other side $N$ and $\tilde{N}$ are the same on both sides of the duality). These states have a mass:
\be
M = \frac{1}{R_S} =\frac{\left(\mathbf{A}_E - \mathbf{\omega}_E\right)^2}{R_E} + R_E
\ee
which gives a relation between the moduli on both sides. Because the mass formula contains the square of the mass, there is a minus sign ambiguity between the identification of the two heterotic descriptions. This is  fixed by choosing an explicit convention.

There is an analogous derivation for the situation in which the labels ``$E$'' and ``$S$'' are interchanged, giving:
\be
\frac{1}{R_E} =\frac{\left(\mathbf{A}_S - \omega_S\right)^2}{R_S} + R_S
\ee

The requirement that these two equations should be compatible with each other leads to
\be \label{wlrel}
\frac{(\mathbf{A}_E -\omega_E)^2}{R_E^2} = \frac{(\mathbf{A}_S -\omega_S)^2}{R_S^2}
\ee

This equation states that the 16-dimensional vectors on the left and on the right are equal, up to an $O(16)$-rotation. At the abstract level, this is the best we can do, as the geometrical relations encoded in the Dynkin diagrams of $E_8 \times E_8$ and $Spin(32)$ only specify their respective root-lattices up to $O(16)$-rotations; moreover, even in an explicit version of the lattice, at either side of the duality $Spin(32)$ or $E_8 \times E_8$-Weyl reflections may be applied to give an equivalent map. The final map can therefore only be fixed by choosing: 1) An explicit realization of the lattices, fixing the freedom for arbitrary $O(16)$-rotations; and 2) An explicit choice for a fundamental domain, fixing the remaining arbitrariness due to the Weyl-group.

In the appendix we have already picked a specific convention, and in the previous, we have chosen fundamental domains (the combination is reflected in the explicit expressions for $\omega_{E,S}$).  In the present conventions, the map between the moduli turns out to take the simple form:
\be
\frac{\mathbf{A}_E -\omega_E}{R_E} = \frac{\mathbf{A}_S -\omega_S}{R_S}
\ee
To verify this formula is not entirely straightforward from the mass formula, because also the expression for the charges changes. The following argument demonstrates the invariance of the mass formula, and therefore that the moduli map is correct.

With the above transformation, the full map on the moduli can be written as:
\be \label{dualmoduli}
(R_E,\mathbf{A}_E-\omega_E) = \left(\frac{R_S}{R_S^2 + (\mathbf{A}_S-\omega_S)^2},\frac{\mathbf{A}_S-\omega_S}{R_S^2 + (\mathbf{A}_S-\omega_S)^2}\right)
\ee
In this form, it is straightforward to see that this can be decomposed into: a shift over $-\omega_S$; an inversion; and a shift over $\omega_E$. 

The shifts correspond to shifts on the charge lattice too. We can use the previous formula's (\ref{shiftsym}), even though the shifts over $\omega_{S,E}$ do not correspond to lattice symmetries. The inversion is the previous ``T-duality'' formula, but we have to correct with a rescaling of the radii with a factor 2, and compensate for a minus sign. The rescaling corresponds to
\be
D_c: \quad \ba{rcl}
\left(\mathbf{q}; n,w \right) & \rightarrow & \left(\mathbf{q};\frac{1}{c} n,2c \right) \\
\left(R,\mathbf{A} \right) & \rightarrow & \left(\frac{1}{c} R, \frac{1}{c} \mathbf{A} \right) \ea
\ee
This scaling was not mentioned before, because it is no symmetry of the charge lattice. This ``non-symmetry'' nevertheless leaves the mass and level-matching formulae invariant. For the symmetry between the two heterotic descriptions we need the version with $c = \hlf$ or $c=2$, depending on whether we wich to rescale before, or after the T-duality transformation. Compensating for the extra minus sign can be done in essentially two ways, we can use a Weyl reflection $W_{-1}$ on $\mathbf{A}$, or we can use the parity transformation $P$ (see eq. (\ref{parsym})). We have remarked before that the product $PW_{-1}$ acts trivially on the moduli, but not on the charge lattice, so we can only decide for $P$ or $W_{-1}$ upon comparing with our conventions for the charges. Checking that the sequence of transformations changes the bases of $\alpha_i$ ($E_8 \times E_8$) to the $\alpha'_i$ of appendix \ref{latcon}, it turns out that we need  the parity symmetry $P$.  

The total transformation from the $E_8 \times E_8$ to $Spin(32)/\Z_2$-string is then given by:
\be
S_{-\omega_S}D_{2}PTS_{\omega_E}
\ee
Because all the intermediate steps that make up the transformation (\ref{dualmoduli}) leave the mass- and level matching formula invariant, the full sequence of transformations, with their simultaneous actions on charge lattice and moduli does too. 

The reader may also wish to compare to explicit examples previously computed in the literature. ($\mathbf{A}_E = \omega_E \leftrightarrow\mathbf{A}_S = \omega_S$ \cite{Ginsparg:1986bx}
, $R_E = \frac{1}{R_S}$; $\mathbf{A}_E=0 \ \rightarrow R_S = R_E/(R_E^2 + 2)$ \cite{Cachazo:2000ey})

The formula's that we have obtained demonstrate that, when expressed in the appropriate variables, the duality from $E_8 \times E_8$-theory to $Spin(32)/\Z_2$ is completely analogous to the map in the other direction, in the variables we have used it amounts to nothing but the interchange of the labels ``S'' and ``E''. The statements in the remainder of this subsection continue to be true under this interchange of labels, so we will give only one of the 2 versions.
 
\FIGURE{
\includegraphics[width=7cm]{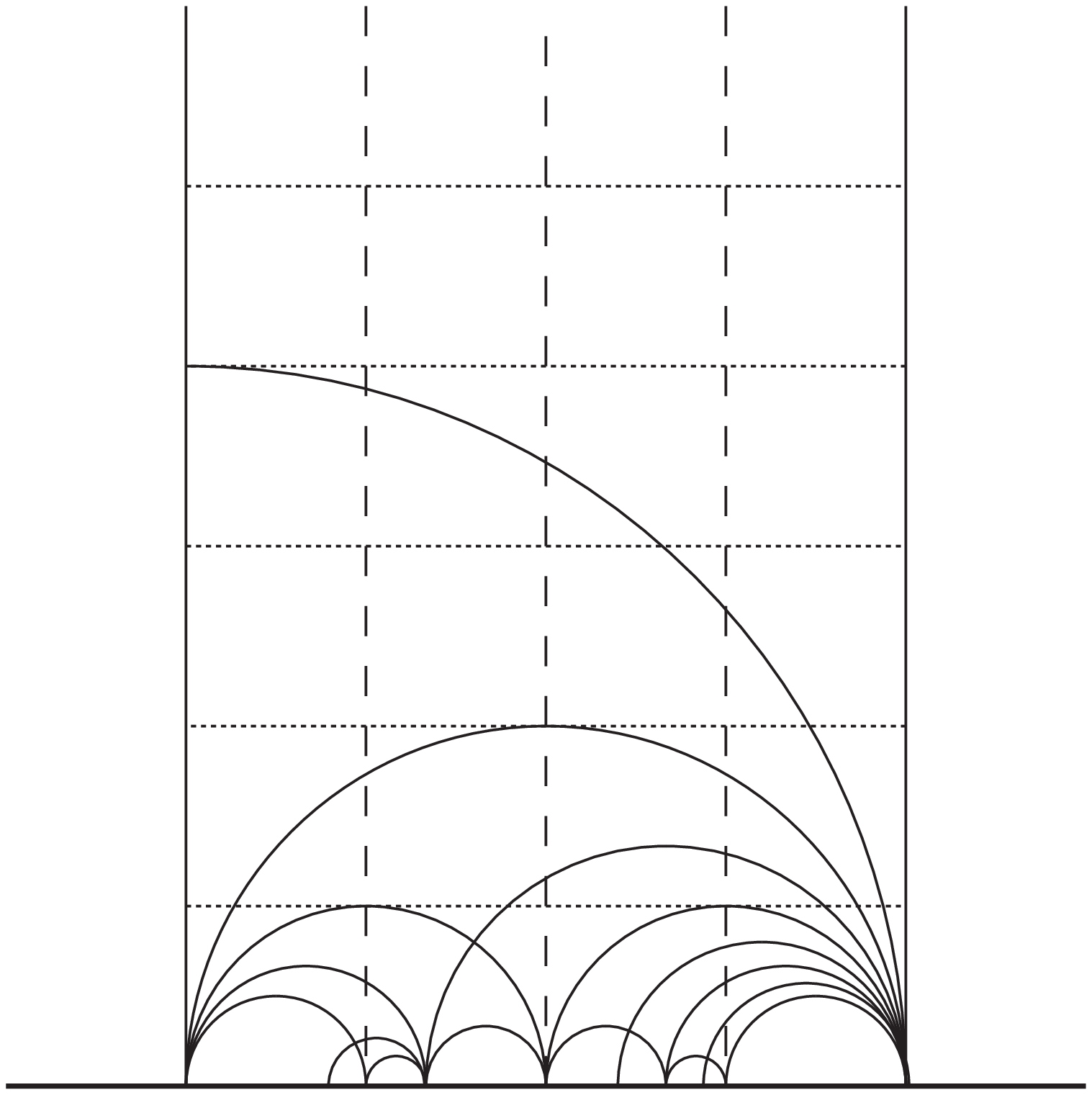}
\includegraphics[width=7cm]{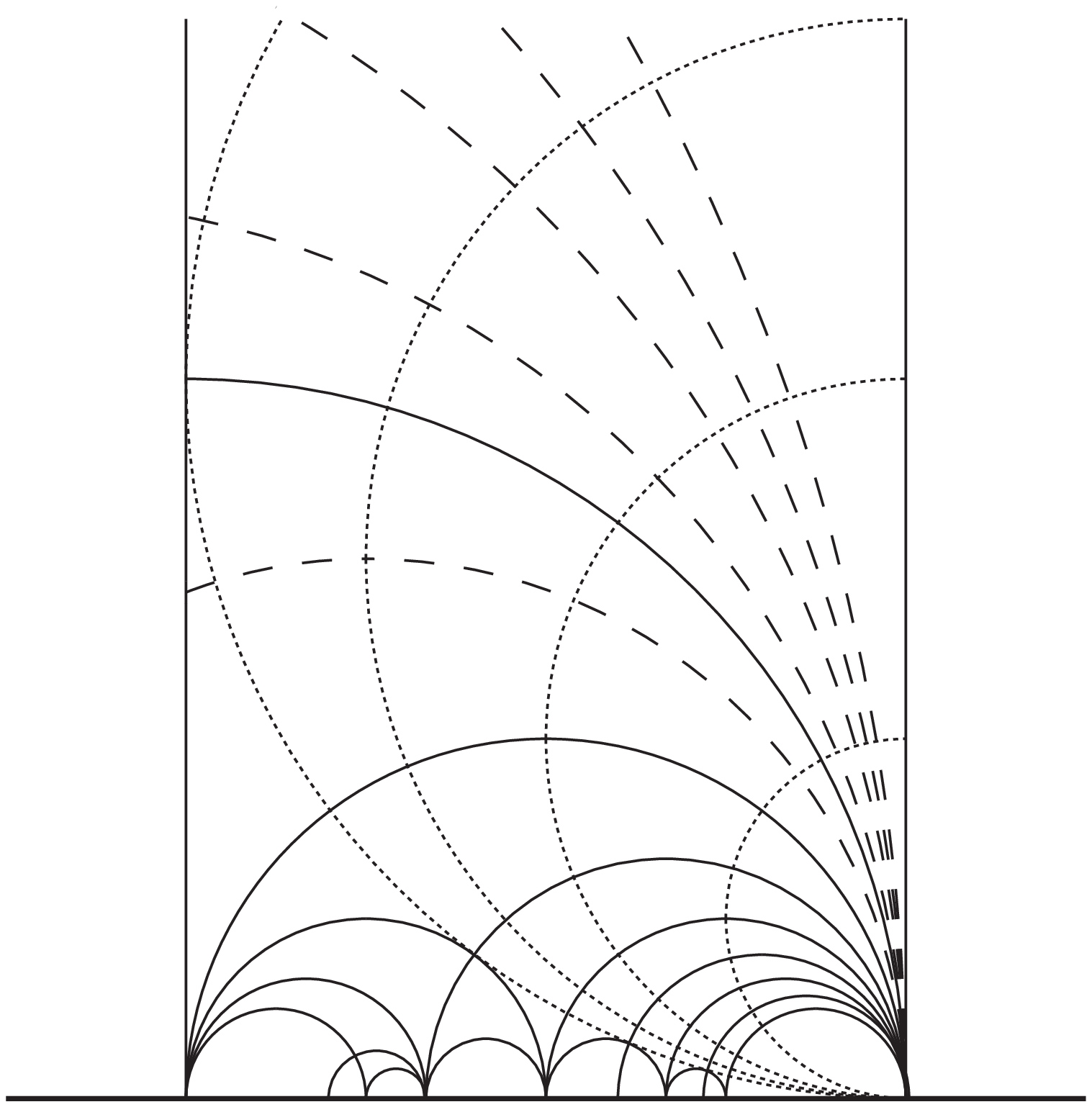}
\caption{Constant radius (small dashes) and constant Wilson line (large dashes) in the two T-dual pictures. More explanations in the text.}\label{hetmodc}
}

A nice way to characterize the hypersurfaces of constant $R_E$ is by the equation:
\be
\left(\mathbf{A}_S -\omega_S \right)^2 + \left(R_S-\frac{1}{2R_E} \right)^2 = \frac{1}{4R_E^2}
\ee
which is the equation for a hypersphere, with midpoint at $(1/2R_E,\omega_S)$, and radius $1/2R_E$. As such, the points $(0,\omega_S)$ and $(1/R_E,\omega_S)$ are contained in the hypershere. The first of these two points corresponds to $Spin(32)/\Z_2$-theory at $R_S = \infty$. This is contained in all surfaces of $R_E$ constant as the limit for infinitely large Wilson line (on the covering space of course). The point $(1/R_E,\omega_S)$ corresponds to $(R_S, \omega_S)$ via the Ginsparg map  \cite{Ginsparg:1986bx}.

For constant Wilson lines $\mathbf{A}_E = \mathbf{C} \neq \omega_E$ the equations imply that $\mathbf{A}_S-\omega_S$ has the same direction as $\mathbf{C}-\omega_E$. Taking the innerproduct of equation (\ref{wlrel}) with $(\mathbf{C}-\omega_E)$ then gives an equation for the lines with $A_E=C$ with varying $R_E$:
\be
\left((\mathbf{A}_S -\omega_S) - \frac{\mathbf{C}-\omega_E}{2(\mathbf{C}-\omega_E)^2} \right)^2 + R_S^2 = \frac{1}{4(\mathbf{C}-\omega_E)^2}
\ee
This gives the equation of a circle, with midpoint at the $R=0$-plane, and radius and midpoint such that the circle passes through $(0,\omega_S)$. As the latter point is a dual image of $R_E \rightarrow \infty$, it should be clear that all lines for constant Wilson line $A_E$ should pass through this point; happily, our map tells us that they do.

For completeness we remind the reader that the well-known case $\mathbf{A}_E = \omega_E$ corresponds to a straight line $\mathbf{A}_S =\omega_S$, with $R$ arbitrary; this can again in the usual way be regarded as the limit of a circle with infinite radius.
 
\subsection{The ``smallest'' radius}

We now have two sets of coordinates on the moduli space, one set of $E_8 \times E_8$-coordinates, and one set of $Spin(32)/\Z_2$ coordinates, and we know how the radii in both picture are related. Standard wisdom tells us that we should preferably use the picture in which the radius is bigger than in the other picture. We should then first determine the surface at which the radii are equal.

Solving for $R_E =R_S=R$ then gives
\be
\left(\mathbf{A}_E - \omega_E\right)^2 + R^2 = 1
\ee
Also, for $R_E = R_S$ we have $\mathbf{A}_E-\omega_E = \mathbf{A}_S -\omega_S$, so the appearance of $E_8$-quantities in this formula is superficial. At $\mathbf{A}_E = \omega_E$ the minimal radius is 1, but by moving away from this point the minimal radius can be made smaller.

This sphere is intersecting with the sphere that bounds the fundamental domain from below
\be
\mathbf{A}_E^2 + R^2 = 2
\ee
At the intersection we have
\be \label{hyp}
\mathbf{A}_E \cdot \omega_E = \thlf
\ee
This determines a surface intersecting with the spheres. The minimum value that the radius can reach is given by maximizing $\mathbf{A}_E^2$, within the fundamental domain, over this surface.

One can again make use of the fact that the fundamental domain is a convex polygon, and so is its intersection with the hypersurface (\ref{hyp}). The maximal Wilson line (maximal $\mathbf{A}^2$) obeying (\ref{hyp}) within the fundamental domain is then (in $E_8 \times E_8$-coordinates)
\be
\mathbf{A}_E = \left(0^4,(-\qrt)^3, \tqrt, -\tqrt, (\qrt)^3, 0^8 \right)
\ee
Or, in $Spin(32)$-coordinates 
\be
\mathbf{A}_S = \left( (\hlf)^4, (\qrt)^8, 0^4 \right)
\ee
Both of these have 
\be
\mathbf{A}_E^2 = \mathbf{A}_S^2 = \thlf
\ee
and hence correspond to the minimal radius
\be
R^2 = \hlf
\ee
At this radius and Wilson line there is enhanced symmetry. The radius is at the surface of critital radii. The $E_8 \times E_8$ Wilson line would leave $SO(10)^2 \times SU(4)^2$ at a generic radius, the $Spin(32)$-version leaves $SO(8)^2 \times SU(8)$ at a generic radius, but at the critical radius both are enhanced to $SO(10)^2 \times SU(8)$ (obtained by deleting the nodes $\alpha_6$ and $\alpha_{14}$ from the diagram for $\Pi_{17,1}$, see appendix \ref{latcon}). As this is a gauge group of rank 17, it is a maximal symmetry group.
 
\subsection{Breaking and recombination of $E_8 \times E_8$} \label{breakrecomb}

Even though we have established that there are many cases where the $R \rightarrow 0$ limit of the $E_8 \times E_8$ string is again an $E_8 \times E_8$ string, this does not mean that everything is ``unchanged''. In the presence of a non-trivial Wilson-line $E_8 \times E_8$ is broken at radii different from $0$ and $\infty$, but interestingly, the way in which the ``pieces'', the unbroken subgroups, are glued together for $R \rightarrow 0$ need not be the same as for $R \rightarrow \infty$.

We will illustrate this with an example. Consider the heterotic $E_8 \times E_8$ string, compactified on a circle of radius $R$, with Wilson line
\be
\mathbf{A}= \left( -\hlf, \hlf, 0^{12}, -\hlf, \hlf \right)
\ee
With the techniques established in the previous section, it is easily verified that this breaks the gauge group to $(E_7 \times SU(2))$ (in \cite{deBoer:2001px} it was shown that the topology of this group is given by additional quotienting by the $\Z_2$ generated by the product of the non-trivial center elements of the $E_7$'s and $SU(2)$'s). Both the $R \rightarrow \infty$ limit and the $R \rightarrow 0$ limit result in decompactification to an $E_8 \times E_8$ theory.

The massless gauge bosons take values in the adjoint representations of $E_7$ and $SU(2)$. 

The set of states with
\be
\mathbf{k}_q^2 = 2, k_w =0 \quad \rightarrow \quad w =0, \mathbf{q}^2 =2
\ee
has mass proportional to $1/R$, and will fill in the required continuum when $R \rightarrow \infty$.

There is also a set of states with 
\be
\mathbf{k}_q^2 = 2, k_n =0 \quad \rightarrow \quad (\mathbf{q} -  w \mathbf{A})^2 = \mathbf{q}^2 + 2nw =2
\ee
States obeying these conditions will form a continuum when $R \rightarrow 0$. That such states exist is not obvious (and in the general case not true as a matter of fact), but in this example it is guaranteed: They can in principle be computed by starting with the momentum states of the T-dual theory, and acting upon these states with the inverse transformations that the algorithm has given us.

It is instructive, to determine the quantum numbers (irreducible representations) that these states form under $SU(2) \times E_7 \times E_7 \times SU(2)$. We find:
\be
\bt{|c|c|}
\hline
Mass & $su(2) \oplus e_7 \oplus e_7 \oplus su(2)$ irreps \\
\hline
$M= 0$ & $\mathbf{(3,1,1,1) \oplus (1,133,1,1) \oplus (1,1,133,1) \oplus (1,1,1,3)}$ \\
$M \propto 1/R$ & $\mathbf{(2,56,1,1) \oplus (1,1,56,2)}$ \\
$M \propto R$  & $\mathbf{(2,1,56,1) \oplus (1,56,1,2)}$ \\
\hline
\et
\ee
Comparing the representation content of the states that become massless in the different limits, one sees that they are essentially the same, but the representation content has been permuted. As a consequence, in the $R\rightarrow 0$ limit, the $SU(2)$'s do not team up with their ``original'' $E_7$-counterparts, but they swith partners. Roughly we have
\bd
\ba{|c|}
\hline
E_8 \\
\hline 
 E_8 \\
\hline
\ea \quad \stackrel{R \rightarrow \infty}{\longleftarrow} \quad
\ba{|cc|}
\hline
E_7 & SU(2)\\
\hline 
SU(2)& E_7 \\
\hline
\ea \quad \leftrightarrow \quad
\ba{|c|c|}
\hline
E_7 & SU(2) \\ 
SU(2) & E_7 \\
\hline
\ea \quad \stackrel{R \rightarrow 0}{\longrightarrow} \quad
\ba{|c|c|}
\hline
E_8 & E_8 \\
\hline
\ea 
\ed
It is obvious that the T-dual theory has the same Wilson line as the original theory. The states with masses scaling as $1/R$ actually occur in multiples of $1/2R$ (this is a consequence of the possible values for $\mathbf{q \cdot A}$), while the states with masses scaling with $R$ have masses that are integer multiples of $R/2$. The exchange of these menas that, if $R$ and $R'$ are the radii of the original respectively the T-dual theory, that they obey
\be
RR' = 1 \qquad (= \alpha'/2)
\ee

In the above example we have broken $E_8 \times E_8$ symmetrically, breaking each $E_8$ to two semi-simple factors (without additional $U(1)$'s). There are more possibilities to do this, each given by using a symmetric combination of fundamental weights as Wilson line. The result is the following table
\be \label{tab1}
\bt{|cc|c|}
\hline
$G$ & $H$ & $\alpha'/RR$ \\
\hline
$E_7$ & $SU(2)$ & 2 \\
$E_6$ & $SU(3)$ & 3 \\
$SO(10)$ & $SU(4)$ & 4 \\
$SU(5)$ & $SU(5)$ & 5 \\
$SU(2) \times SU(3)$ & $SU(6)$ & 6 \\
\hline
\et
\ee
All the groups in this table break up and recombine under T-duality, along the pattern
\bd
\ba{|c|}
\hline
E_8 \\
\hline 
 E_8 \\
\hline
\ea \quad \stackrel{R \rightarrow \infty}{\longleftarrow} \quad
\ba{|cc|}
\hline
G & H\\
\hline 
H & G \\
\hline
\ea \quad \leftrightarrow \quad
\ba{|c|c|}
\hline
G & H \\ 
H & G \\
\hline
\ea \quad \stackrel{R \rightarrow 0}{\longrightarrow} \quad
\ba{|c|c|}
\hline
E_8 & E_8 \\
\hline
\ea 
\ed
We have also mentioned the relation between radius and dual radius in table (\ref{tab1}). 

We note in passing that the above table gives examples for the description of heterotic T-duality in \cite{deBoer:2001px}, appendix B.1., for values of $k=2,3,4,5,6$. Also, this computation provides us with further evidence for the fact that the $\Z_2$ automorphism exchanging the two $E_8$'s must be regarded as a local symmetry of heterotic $E_8 \times E_8$ strng theory, as it reveals that any assignment of ``first'' and ``second'' $E_8$ is entirely ambiguous.

Two more symmetric combinations of fundamental coweights break the symmetry to $(SU(8) \times SU(2))^2$ and $SU(9)^2$ respectively. These have $E_8 \times E_8$ duals, but not the recombination phenomenon.

The last symmetric combination of fundamental weights breaks $E_8 \times E_8$ to $SO(16) \times SO(16)$; it is of course well known that these Wilson lines allow for a very special kind of recombination (to $Spin(32)$). 

\section{Discussion, conclusions}

We have studied in some detail the $R \rightarrow 0$ limit for heterotic strings on a circle. Though, except for a finite number of points, the $R\rightarrow 0$ limit takes us out of (any choice for) the fundamental domain, there are a number of reasons why such an analysis is nevertheless usefull.

An important justification is that physical intuition takes place on the covering space; it is common to loosely speak of taking large or small volume limits, without bothering (a priori) whether this takes us out of
fundamental domains. It is common lore that string theory should avoid ``small volume'' singularities, resolving them in some way. T-duality is one of the most cherished mechanisms for achieving this; one of the aims of the present paper was however to demonstrate that even in fairly simple theories (heterotic strings on a circle) T-duality acts more subtle than a simple map back to ``large volume''. This is already implicit in the realization that geodesic motion on string moduli spaces generically gives rise to chaotic behaviour \cite{Horne:1994mi}, but the present analysis serves to emphasize this and make it explicit.

In the past the chaotic properties of motion on string moduli spaces have been proposed for selection mechanisms in cosmological models \cite{Horne:1994mi} (similar mathematics plays a role in \cite{Damour:2000hv} though the physics of these models is slightly different). Also for the heterotic string on a circle, similar remarks can be made; though we have focussed on the subset of geodesics with regular limits, it is simple to see that for a generic geodesic neither of the two endpoints corresponds to a decompactification limit. Cosmological models based on heterotic strings on a circle however are far from realistic, and can be no more than a toy model displaying behaviour that may be qualitatively similar to more realistic models. 

Another motivation for the present study is to get more handles on string dualities, especially for (asymmetric) orbifolds of the heterotic string. Such models are known to exhibit interesting dualities (some examples can be found in \cite{deBoer:2001px, Lerche:1997rr, Witten:1997bs}). To check such dualities knowledge of the large volume limits in the heterotic moduli space is indispensable (see also \cite{Mikhailov:1998si}). As an example, the special Wilson lines leading to the construction of the ``triples'' of \cite{deBoer:2001px}, are conjugate to the Wilson lines studied in section \ref{breakrecomb}, providing us with a number of T-dualities for these models. We hope to report on further applications in the future.

\acknowledgments

Several discussions with Jarah Evslin on T-duality for heterotic strings served as inspiration for the work presented here. I would also like to thank Marc Henneaux, and Daniel Persson for a discussion on chaos, and for bringing reference \cite{Horne:1994mi} to my attention, and the theoretical physics group in Groningen for hospitality and comments and questions during a stage of this work. This work was supported in part by the ``FWO-Vlaanderen'' through project G.0034.02, in part by the Belgian Federal Science Policy Office through the Interuniversity Attraction Pole P5/27 and in part by the European Commission RTN programme HPRN-CT-2000-00131, in which the author is associated to the University of Leuven.

\newpage

\appendix
\section{Useful conventions for gauge theories on a circle}
\subsection{$E_8$-gauge theory} \label{appe8}

\FIGURE{
\includegraphics[width=6cm]{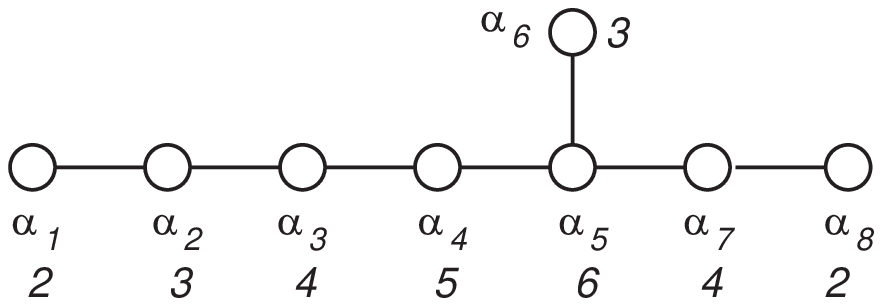}
\caption{The Dynkin diagram of $E_8$.}\label{E8}
}

The Dynkin diagram for $E_8$ is given in figure \ref{E8}. For the discussion on the moduli-space of $E_8$ flat connections one needs the simple roots of $E_8$, which can be taken to be
\bea
\beta_i & = & e_{i+1} -e_{i+2}\qquad i \in \{1,\ldots,6 \} \\
\beta_7 & = & e_7 +e_8              \\
\beta_8 & = & -\sum_{n=1}^8 e_i  \qquad 3 \leq i \leq 8     
\eea
It is essentially in this form that they appear in the Dynkin diagram of $\Pi_{17,1}$ in $\alpha_1$ to $\alpha_8$ (with extra sign, and the order of the components reversed), and in $\alpha_{12}$ to $\alpha_{19}$.

The highest root in these conventions is given by 
\be
\beta_H =2 \beta_1 + 3 \beta_2 + 4 \beta_3 + 5 \beta_4 + 6 \beta_5 + 3 \beta_6 + 4 \beta_7 + 2 \beta_8 = -e_1 +e_2
\ee
(and occurs in the $\mathbf{q}$-part of $\alpha_9$ and $\alpha_{11}$), and hence the (co)root integers are
\be
a_0 = 1; a_1 = 2; a_2 = 3; a_3 = 4; a_4 = 5; a_5 = 6; a_6 = 3; a_7 = 4; a_8 = 2;
\ee
The reader will notice that these coefficients also play a role in the linear dependency relation for the roots that make up the Dynkin diagram of $\Pi_{17,1}$ (see appendix \ref{latcon}.
 
The coweights in these conventions, defined by $\inp{\beta_i}{\omega_j} = \delta_{ij}$, are
\bea
\omega_i & = & -ie_1 +\sum_{n=2}^{i+1} e_i \qquad i \in \{1, \ldots,5 \}  \\
\omega_6 & = & -\frac{5}{2} e_1 + \hlf (e_2 +e_3 + e_4+e_5+e_6+ e_7 - e_8)\\
\omega_7 & = & -\frac{7}{2} e_1 + \hlf (e_2 +e_3 + e_4+e_5+e_6+ e_7 + e_8)\\
\omega_8 & = & 2e_1   
\eea

\subsection{$Spin(32)/\Z_2$} \label{appspin}

\FIGURE{
\includegraphics[width=13cm]{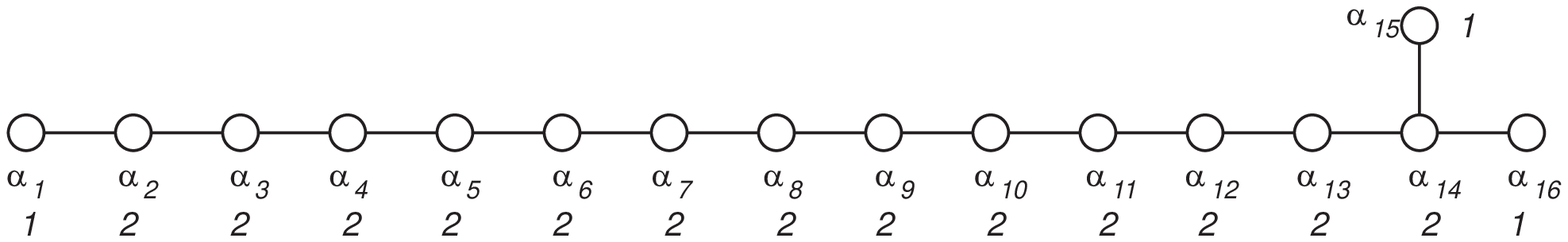}
\caption{The Dynkin diagram of $\Pi_{17,1}$. The enumeration of the nodes and the coefficients apearing in the linear dependence relation are displayed.}\label{D16}
}

Figure \ref{D16} shows the Dynkin diagram for $Spin(32)$ (also known as $D_{16}$).
The simple roots of $Spin(32)$, can be taken to be
\bea
\beta'_i & = & e_{i} -e_{i+1}\qquad i \in \{1,\ldots,15 \} \\
\beta'_{16} & = & e_{15} + e_{16}            
\eea
In this form they appear in the Dynkin diagram of $\Pi_{17,1}$ in $\alpha'_3$ to $\alpha'_{18}$ (see appendix \ref{latcon}).

The highest root is in these conventions given by 
\be
\beta_H =\beta_1 + \beta_{15} + \beta_{16} + 2 \sum_{n=2}^{14}  \beta_n = e_1 + e_2
\ee
(and occurs in the $\mathbf{q}$-part of $\alpha'_2$), and hence the (co)root integers are
\be
a_0 = a_1 =a_{15} = a_{16} =1; a_i  = 2 \qquad 2 \leq i \leq 14.
\ee
These coefficients do not appear in the linear dependency relation for the roots that make up the Dynkn diagram of $\Pi_{17,1}$, but there is another relation that we explain in the text.
 
The coweights in these conventions, defined by $\inp{\beta_i}{\omega_j} = \delta_{ij}$, are
\bea
\omega_i & = & \sum_{n=1}^i e_n \qquad i < 15  \\
\omega_{15} & = & \hlf( \sum_{n=1}^{15} e_n - e_{16})\\
\omega_{16} & = & \hlf \sum_{n=1}^{16} e_n
\eea

\section{Conventions for $\Pi_{17,1}$} \label{latcon}
 
The lattice of charges is encoded in the following Dynkin diagram.

\FIGURE{
\includegraphics[width=13cm]{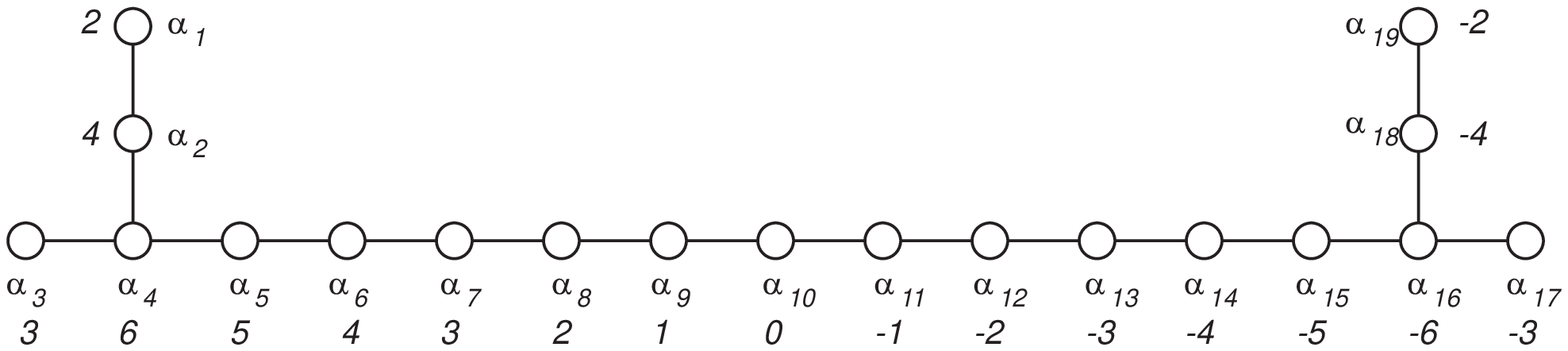}
\caption{The Dynkin diagram of $\Pi_{17,1}$. The enumeration of the nodes and the coefficients apearing in the linear dependence relation are displayed.}\label{Pi17}
}

The Dynkin diagram provides only an abstract encoding of the geometrical relations by the charge vectors generating the charge lattice. The realization of these requires an innerproduct with Lorentzian signature.

Two realizations of this basis of simple roots are exhibited. We will write the corresponding charge-vectors in the format $(\mathbf{q};n,w)$,
with norm (which, in the standard way also defines the inner product)
\bd
\mathbf{q}^2 + 2nw
\ed
The vectors $\mathbf{q}$ take values in $\R^{16}$, and correspond to (and weight) vectors of the ``manifest'' gauge group. We denote by $e_n$ ($n \in \{1, \ldots, 16\}$) a basis of orthonormal unit vectors for $\R^{16}$.

The $E_8 \times E_8$ charge vectors are given by (for  $i \in \{3,4,5,6,7,8,12,13,14,15,16,17 \}$ the charge vectors have a similar structure)
\be
\ba{c@{= (}c@{;}r@{,}r@{)}}
\alpha_1 & \hlf \sum_{i=n}^8 e_n & 0 & 0  \\
\alpha_2 & -e_1 -e_2             & 0 & 0  \\
\alpha_i & e_{i-2} - e_{i-1}     & 0 & 0 \\
\alpha_9 & e_{7}- e_{8}          & 1 & 0 \\
\alpha_{10} & 0                  & -1 & -1 \\
\alpha_{11} & e_9 -e_{10}        & 1 & 0 \\
\alpha_{18} & e_{15}+e_{16}      & 0 & 0 \\
\alpha_{19} & -\hlf \sum_{n=9}^{16} e_n & 0 & 0 \\
\ea
\ee 

A second realization makes manifest the $D_{16}$ (extended) Dynkin-diagram hidden in $\Pi_{17,1}$ manifest. We add a prime to all the $\alpha'_i$ to facilitate the discussion on comparing the two realizations of the basis, and now $i$ runs from $3$ to $17$:
\be
\ba{c@{= (}c@{;}r@{,}r@{)}}
\alpha'_1 & 0                     & -1 & -1  \\
\alpha'_2 & -e_1 -e_2             & 1 & 0  \\
\alpha'_i & e_{i-2} - e_{i-1}     & 0 & 0 \\
\alpha'_{18} & e_{15}+e_{16}      & 0 & 0 \\
\alpha'_{19} & -\hlf \sum_{n=1}^{16} e_n & 1 & -1 \\
\ea
\ee 

Among the roots encoded by the nodes of this Dynkin diagram there is one linear relation (so in a sense, it is not really appropriate to call it a Dynkin diagram, as ususally linear independence among the roots is a requirement entering the definitions.). The linear relation takes the form
\bea
 2 \alpha_1 + 4 \alpha_2 + 3 \alpha_3 + 6 \alpha_4 + 5 \alpha_5
+ 4 \alpha_6 + 3 \alpha_7 + 2 \alpha_8 + \alpha_9 & &  \non \\
 -\alpha_{11}-2 \alpha_{12} - 3\alpha_{13} - 4 \alpha_{14} -5 \alpha_{15} - 6 \alpha_{16} - 3 \alpha_{17} - 4 \alpha_{18} -2 \alpha_{19}& = & 0 \label{lindep}
\eea
(and of course, the same relation applies when the $\alpha_i$ are replaced by the $\alpha'_i$).

The diagram \ref{Pi17} has a rather obvious $\Z_2$ reflection symmetry. It is obvious how this symmetry acts in $E_8 \times E_8$ coordinates: AS an interchange of the 2 $E_8$ sub-diagrams. In $Spin(32)$-coordinates the symmetry acts as
\be
(\mathbf{q}, n, w) \rightarrow (-\mathbf{q}^R + w \left( (\hlf)^{16} \right), n + \mathbf{q} \cdot \left((\hlf)^{16}\right) - 2w, w)
\ee
This is the natural extension to heterotic string theory of the symmetry
$\mathbf{q} \rightarrow -\mathbf{q}^R$ of the (extended) $Spin(32)$ Dynkin diagram. In the description of the moduli space of section PWAP, we need the shift over $\left( (\hlf)^{16} \right)$ to map the fundamental domain back to the fundamental domain. Obviously, this $\Z_2$ symmetry leaves the linear dependence relation (\ref{lindep}) invariant (up to an irrelevant overall sign).

In the derivation of T-duality of the two heterotic theories, use is made of a state with $n=1$ unit of momentum, and all other charges set to zero. Such a state is formed in the $E_8 \times E_8$ basis as:
\be
\Omega_{E,\alpha} = 2 \alpha_1 + 4 \alpha_2 + 3 \alpha_3 + 6 \alpha_4 + 5 \alpha_5
+ 4 \alpha_6 + 3 \alpha_7 + 2 \alpha_8 + \alpha_9 = (0;1,0) 
\ee
and consequently, corresponds in the $Spin(32)/\Z_2$ basis to
\bea
\Omega_{E,\alpha'}  & = & 2 \alpha'_1 + 4 \alpha'_2 + 3 \alpha'_3 + 6 \alpha'_4 + 5 \alpha'_5
+ 4 \alpha'_6 + 3 \alpha'_7 + 2 \alpha'_8 + \alpha'_9 \non \\
&  = & (-\sum_{i=1}^8 e_i;2,-2) = (-2 \omega_S;2,-2)
\eea

The state with a single unit of momentum in the $Spin(32)$-basis is given by
\be
\Omega_{S,\alpha'} = \alpha'_2 + \alpha'_3 + \alpha'_{17} + \alpha'_{18} + 2 \sum_{n=4}^{16} \alpha'_n = (0;1,0) 
\ee
and corresponds in the $E_8 \times E_8$-picture to
\bea
\Omega_{S,\alpha} &  = &  \alpha_2 + \alpha_3 + \alpha_{17} + \alpha_{18} + 2 \sum_{n=4}^{16} \alpha_n \non \\
& =  & (2(e_9-e_8);2,-2) = (-2 \omega_E;2,-2) 
\eea
The objects $\omega_S$ and $\omega_E$ that are crucial in the analysis of Wilson lines appear naturally in this description, and the above way of defining them allows for an easy translation to other conventions for the $E_8 \times E_8$ or $Spin(32)/\Z_2$ lattices.

\end{document}